\documentclass[preprint]{ptephy_v1}

\usepackage{hyperref}

\usepackage{bm}
\usepackage{ulem}

\begin{document}

\title{Global density-dependent $\alpha$-nucleon interaction for $\alpha$-nucleus elastic scattering}


\author{T. Furumoto}
\affil{College of Education, Yokohama National University, Yokohama 240-8501, Japan \email{furumoto-takenori-py@ynu.ac.jp}}

\author{K. Tsubakihara}
\affil{National Institute of Technology, Asahikawa College, Asahikawa 071-8142, Japan}

\author{S. Ebata}
\affil{Graduate School of Science and Engineering, Saitama University, Saitama 338-8570, Japan}

\author{W. Horiuchi}
\affil{Department of Physics, Osaka Metropolitan University, Osaka, 558-8585, Japan}
\affil{Nambu Yoichiro Institute of Theoretical and Experimental Physics (NITEP), Osaka Metropolitan University, Osaka 558-8585, Japan}
\affil{RIKEN Nishina Center, Wako 351-0198, Japan}
\affil{Department of Physics, Hokkaido University, Sapporo 060-0810, Japan}


\begin{abstract}%
We provide a global density-dependent $^4$He-nucleon (DD-$\alpha N$) interaction 
to construct the $\alpha$-nucleus optical model potential (OMP) in a wide range of incident energies.
The global parametrization for the DD-$\alpha N$ interaction is obtained based on the proton-$^4$He OMP which reproduces the elastic scattering cross-section data very well in the incident energies of 12.04--500 MeV per nucleon.
We derive the $\alpha$-nucleus potential by a folding procedure with the point-nucleon density obtained by a microscopic mean-field model using the present DD-$\alpha N$ interaction.
The density dependence of the DD-$\alpha N$ interaction is fixed phenomenologically to reproduce the $\alpha$-nucleus elastic scattering cross-section data by the $^{16}$O, $^{40}$Ca, $^{58}$Ni, $^{90}$Zr, and $^{208}$Pb targets at $E/A =$ 10--342.5 MeV.
We also show the total reaction cross sections, which are helpful in fixing one free parameter, the renormalization factor for the imaginary part of the $\alpha$-nucleus potential.
Lastly, we show some examples, which clearly demonstrate the validity and power of the present DD-$\alpha N$ approach.
\end{abstract}

\subjectindex{D06, D22}

\maketitle

\section{Introduction}
A microscopic description of the optical model potential (OMP) is useful for describing elastic scattering phenomena in a non-empirical way.
The nucleus-nucleus OMP has often been constructed by the double-folding model, where the densities of the interacting nuclei are folded by the nucleon-nucleon interaction~\cite{SIN75, SAT79}.
Many efforts have been devoted to obtaining a reliable OMP by improving the folding prescriptions and the effective nucleon-nucleon interaction~\cite{SIN75, SAT79, KOB82, DNR, KHO94, SAT97, KHO97, KHO01, FUR06, KHO07, FUR09, FUR10, FUR16}.
Especially, the effect of the exchange term, the density-dependence of the effective nucleon-nucleon interaction, and the treatment of the local-density approximation have been intensively studied:
The suitable treatment of the finite-range interaction in the exchange term was proposed in Refs.~\cite{KHO94, KHO00, KAT02}.
The density dependence of the nucleon-nucleon interaction was introduced to incorporate the properties of the nuclear matter~\cite{PET75, SAT79, KOB82, KHO95, KHO97, FUR09}.
The treatment of the local-density approximation for the nucleus-nucleus system has been discussed~\cite{FUR06, KHO07, FUR16}.
From these findings, the frozen-density approximation can be a plausible prescription for describing heavy-ion scattering with the density-dependent nucleon-nucleon interaction satisfying the saturation property in the nuclear medium.

The $\alpha$ particle ($^4$He nucleus) plays an important role in nuclei and its scattering is a major reaction in the astrophysical environment~\cite{NRA}.
Thus, plenty of the microscopic $\alpha$-nucleus potential is proposed in the folding model approaches~\cite{CAR96, SAT97, KHO97, KHO01, FUR06, KHO07, EGA14, TOY15-2, TOY18}.
However, the treatment of the local density in the construction of the $\alpha$-nucleus folding model potential (FMP) is not consistent, which is in contrast to the nucleus-nucleus OMP for heavy-ion scattering.
As we will see in the following examples, the $\alpha$-nucleus potential is somewhat heterogeneous compared to the nucleus-nucleus potential.

The CDM3Y6~\cite{KHO97} and JLM~\cite{JLM77} interactions have often been used to construct the FMPs.
It is known that the double-folding model potentials with the CDM3Y6 and JLM interactions need some corrections to reproduce the elastic scattering cross-section data.
The so-called renormalization factors are often introduced as
\begin{equation}
U=N_R V_{\rm CDM3Y6} + iW_{\rm phenome.}
\end{equation}
and
\begin{equation}
U=N_R V_{\rm JLM} + i N_I W_{\rm JLM}
\end{equation}
for the CDM3Y6 and JLM interactions, respectively.
Here, $V_{\rm CDM3Y6}$ is the FMP with the CDM3Y6 interaction.
$V_{\rm JLM}$ and $W_{\rm JLM}$ are the real and imaginary parts of the FMP with the JLM interaction, respectively.
$W_{\rm phenome.}$ is the phenomenological imaginary potential.
The $N_R$ and $N_I$ are the renormalization factors for the real and imaginary parts of the potential $U$, respectively.
To clarify the problem in constructing the $\alpha$-nucleus potential, we compare $N_R$ for the $\alpha$-nucleus and nucleus-nucleus cases.
\begin{figure}[h]
\centering
\includegraphics[width=6.4cm]{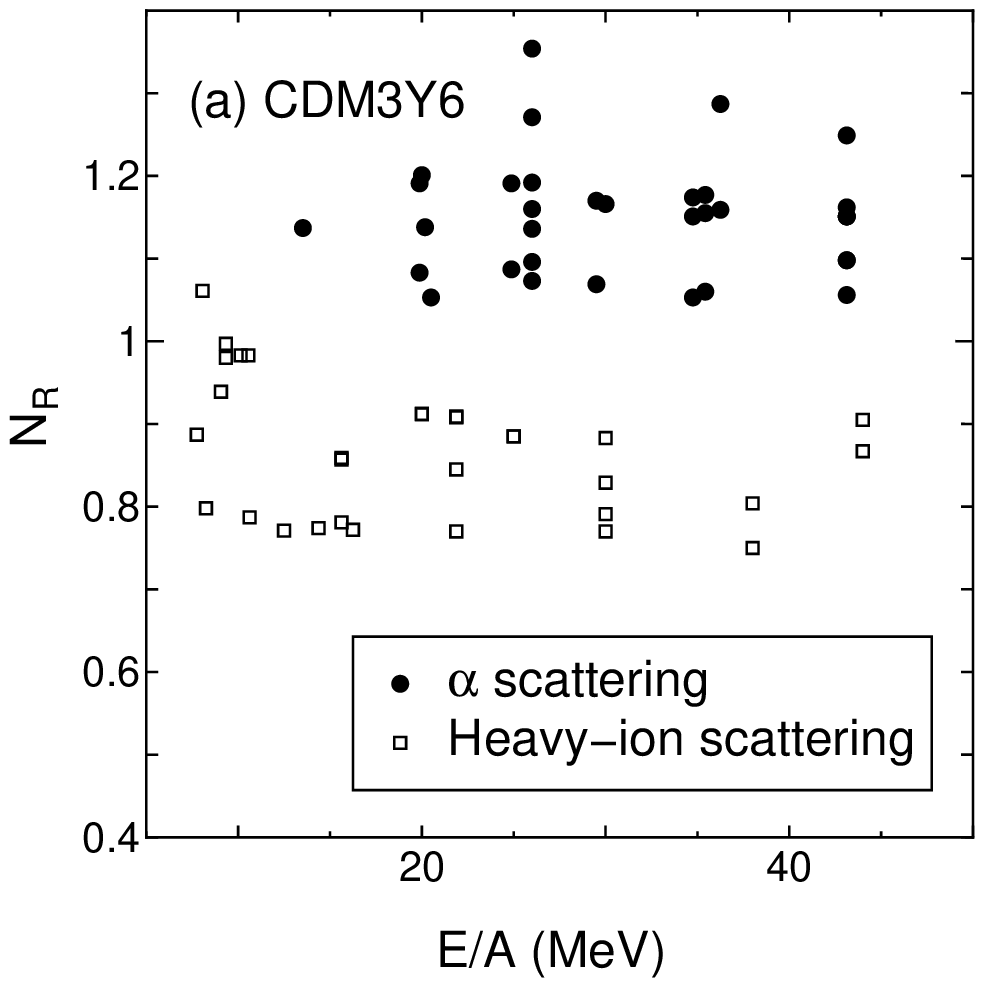}
\hspace{10mm}
\includegraphics[width=6.4cm]{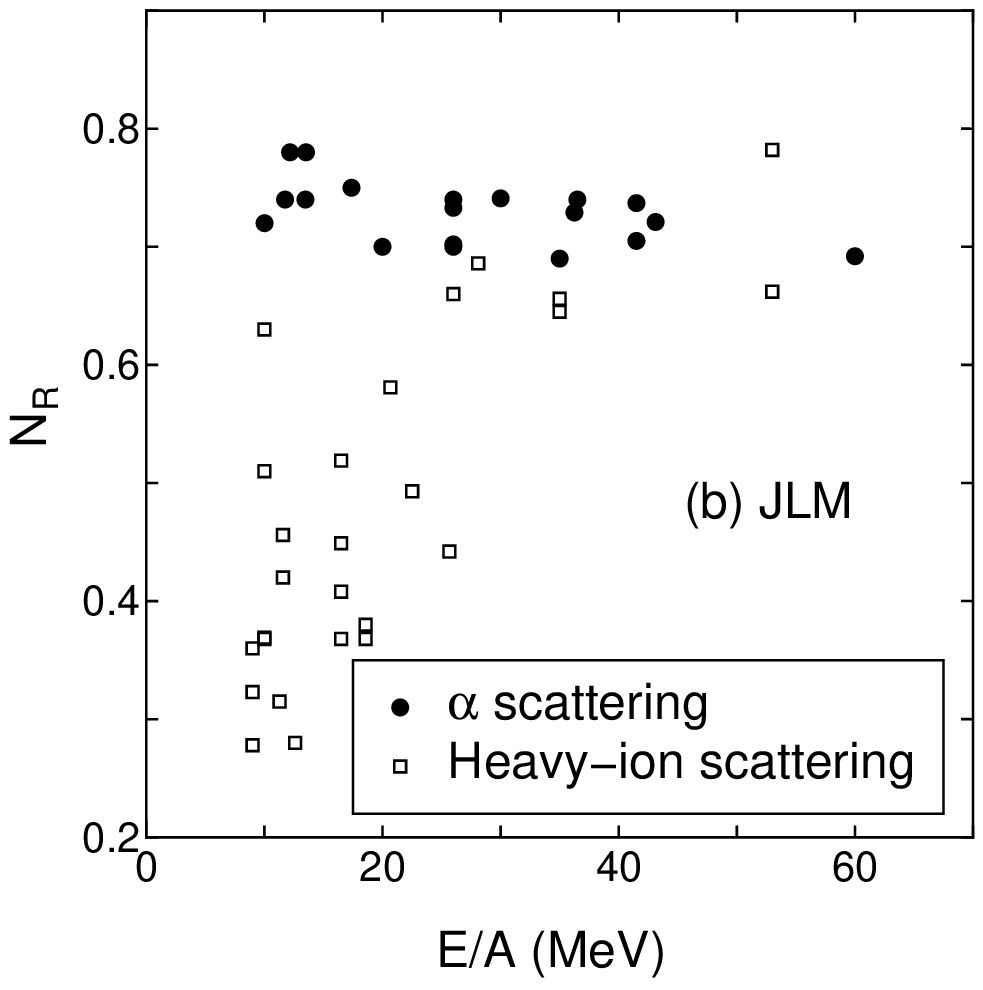}
\caption{Renormalization factors for the real part of the FMP obtained from the (a) CDM3Y6 and (b) JLM interactions.
$E/A$ is the incident energy per nucleon.
The theoretical data are taken from Refs.~\cite{KHO97, KHO00, KHO00-2, OGL00, KHO01, KHO05, CAR96, TRA00, FUR06}.}
\label{fig:nr}
\end{figure}
Figure~\ref{fig:nr} plots $N_R$ for the CDM3Y6 and JLM interactions, respectively.
The filled circles and open squares indicate the values for the $\alpha$-nucleus and nucleus-nucleus systems, respectively.
Here, the nucleus-nucleus system means that the mass number of the incident particle is greater than 6, i.e., the heavy-ion scattering.
The mass of the target nucleus is widely employed $A = $ 12--208.
We note that the $N_R$ values with the CDM3Y6 interaction for the nucleon elastic scattering also give similar results for the heavy-ion scattering.
The $N_R$ values are 0.901--0.990~\cite{KHO02}.
On the other hand, the $N_R$ values with the JLM interaction for the nucleon elastic scattering are not smoothly connected to both the $\alpha$ and heavy-ion scatterings by the treatment of the local density approximation.
Therefore, the $N_R$ values with the JLM interaction for the nucleon elastic scattering are not discussed here.
It appears that the trend of the $N_R$ values for the $\alpha$-nucleus system is apparently different from that for the nucleus-nucleus system.
This anomaly occurs when the same prescription in the folding procedure is applied for the $\alpha$-nucleus and nucleus-nucleus systems.
We remark that recently the target density approximation (TDA) was proposed to reproduce the $\alpha$-nucleus scattering cross-section data~\cite{EGA14, TOY15}.
However, the TDA totally ignores the nuclear medium (density) effect from the $\alpha$ particle.
Furthermore, in the TDA, the description of the $\alpha$-nucleus potential is inconsistent with that of the nucleus-nucleus potential.
The anomaly of the $\alpha$-nucleus potential remains unresolved.

In this paper, we construct a density-dependent $\alpha$-nucleon (DD-$\alpha N$) interaction to address the problem.
In general, the effective nucleon-nucleon interaction for the nuclear reaction is not designed for the $\alpha$-nucleus FMP.
For example, the density dependence of the CDM3Y6 interaction and most of the effective interactions is fixed to satisfy the saturation property of the infinite nuclear matter.
It should be noted that the $\alpha$ particle is far from the condition of the infinite nuclear matter:
The inner part of the density for the $\alpha$ particle is substantially larger than the saturation density, and the binding energy of the $\alpha$ particle is not explained by the mass formula that accounts for the saturation energy in the infinite nuclear matter.
Therefore, in this paper, we treat the $\alpha$-nucleon system as an elementary process to describe the $\alpha$-nucleus system, not the nucleon-nucleon system.
The density dependence of the $\alpha$-nucleon interaction is fixed to reproduce the experimental data, incorporating the complicated medium effects in a phenomenological way.
Here, we refer to the earlier works for the (DD-)$\alpha N$ interactions~\cite{BER80, SAT97, KOL00}. 
To reproduce the $\alpha$-nucleus scattering, their (DD-)$\alpha N$ interaction includes some free parameters, leading to the inconsistency in the elementary $p$ + $^{4}$He scattering.
Therefore, we propose a new DD-$\alpha N$ interaction that describes the $p$ + $^4$He elastic scattering for constructing a reliable $\alpha$-nucleus FMP for a wide range of incident energies.

The paper is organized as follows.
We first introduce the theoretical framework in Sec.~\ref{sec:formalism}.
The $p$ + $^4$He OMP, the functional forms of the DD-$\alpha N$ interaction, the single-folding model to construct the $\alpha$-nucleus potential, and the input density are explained.
In Sec.~\ref{sec:results}, we show the results of the $p$ + $^4$He elastic scattering.
Then, we present the folding model calculations for the $\alpha$-nucleus elastic scattering for selected target nuclei $^{16}$O, $^{40}$Ca, $^{58}$Ni, $^{90}$Zr, and $^{208}$Pb.
To verify the present OMP, the total reaction cross sections are also calculated and compared with available experimental data.
We lastly show the applications to the $\alpha$ elastic scattering with the present DD-$\alpha N$ interaction.
Finally, we discuss the property and validity of the present DD-$\alpha N$ interaction.
The summary of this paper is given in Sec.~\ref{sec:summary}.

\section{Formalism}
\label{sec:formalism}

In this paper, we construct the $\alpha$-nucleus potential with the newly developed DD-$\alpha N$ interaction using point-nucleon density obtained by the Hartree Fock +BCS (HF+BCS) calculation in the single-folding procedure.
In this section, we first describe the $p$ + $^4$He potential as the elementary process for the $\alpha$-nucleus scattering.
Next, we introduce the functional form of the density dependence of the present $\alpha$-nucleon interaction.
After that, the folding procedure with the DD-$\alpha N$ interaction is explained.

\subsection{$p$ + ${}^4$He potential}

Here, we first construct a phenomenological OMP for a $p$ + $^4$He system.
The form of the OMP, $U$, is assumed by
\begin{eqnarray}
U(s, E/A) &=& f(s; V_0, r_0, a_0) + f(s; V_1, r_1, a_1) + i f(s; W, r_W, a_W) \nonumber \\
&& + \frac{\lambda_{\pi}^2}{r_{\rm SO} 4^{1/3}} \frac{d}{ds} f(s; V_{\rm SO}, r_{\rm SO}, a_{\rm SO}) \ \bm{\ell} \cdot \bm{\sigma} \label{eq:p+4he}
\end{eqnarray}
with
\begin{equation}
f(s; V, r, a) = -\frac{V}{1+\exp{\left( \frac{s-r 4^{1/3}}{a}\right)}},
\end{equation}
where $s$ is the distance between the proton and the $^4$He nucleus.
$E/A$ denotes the incident energy per nucleon.
To adapt the drastic incident-energy dependence of the potential form factor, we employ a double-Woods-Saxon (WS) type potential for the real part of $U$, which can express the wine-bottle shape and repulsive potential at high incident energies~\cite{HAM90, SEN87, FUR10}.
The imaginary part is assumed to be a standard WS type.
For the spin-orbit potential, we also take a standard form called the derivative WS type.
We omit the surface imaginary potential and the imaginary spin-orbit potential because their effects are found to be minor in describing the $p$ + $^4$He elastic scattering.
Though the spin-orbit term vanishes in the single-folding procedure to construct the $\alpha$-nucleus potential, this term plays an important role to describe the $p$ + $^4$He scattering.
Here, we take $\lambda_{\pi}^2 = 2.0$ fm$^2$.
The parameters, $V_0, r_0, a_0, V_1, r_1, a_1, W, r_W, a_W, V_{\rm SO}, r_{\rm SO}, a_{\rm SO}$, are fixed to reproduce the experimental $p$ + $^4$He elastic-scattering cross-section data for each incident energy.
Hereafter, we omit $E/A$ for simplicity.
Their functional form and parameters will be presented in the following section.

\subsection{Global density-dependent $\alpha$-nucleon interaction}

Next, we introduce the density dependence (medium effect) into the $p$ + $^4$He OMP.
The present DD-$\alpha N$ interaction, $U_{{\rm DD-}\alpha N}$, has a following form:
\begin{eqnarray}
U_{{\rm DD-}\alpha N}(s, \rho, E/A) &=& f(s; V_0, r_0, a_0) g_0( \rho, E/A) + f(s; V_1, r_1, a_1) g_1( \rho) \nonumber \\
&& + i f(s; W, r_W, a_W) g_W( \rho), \label{eq:ddan}
\end{eqnarray}
where $g_0$, $g_1$, and $g_W$ are responsible for the density-dependence of the present $\alpha N$ interaction.
Their input density value $\rho$ should be evaluated appropriately, which will be explained later.
The explicit forms of these density-dependent functions are
\begin{eqnarray}
g_0( \rho, E/A) &=& \exp{(- \beta_0 \rho)}, \label{eq:bd-real1} \\
g_1( \rho) &=& 1 + \beta_1 \rho, \label{eq:bd-real2} \\
g_W( \rho) &=& \exp{(- \beta_W \rho)}, \label{eq:bd-imag}
\end{eqnarray}
where $\beta_0$, $\beta_1$, and $\beta_W$ are fixed to reproduce the $\alpha$-nucleus elastic scattering cross-section data.
We note that only $\beta_0$ has the incident energy dependence, $\beta_0 = \beta_0(E/A)$, 
which will be discussed later.
Note that Eqs.~(\ref{eq:bd-real1}) and (\ref{eq:bd-imag}) are always positive and become zero when $\rho$ is high for ensuring the properties of one-pion exchange interaction for the real part and for suppressing the elastic flux inflow for the imaginary part.
We tested some versions of the exponential function as $\exp(-\beta \rho^n)$, and found that the form with $n=1$ is the best reproducing the data.
The core part of Eq.~(\ref{eq:bd-real2}) is referred to Ref.~\cite{SAT97}.
Note that the Eq.~(\ref{eq:ddan}) becomes the bare $\alpha$-nucleon interaction of Eq.~(\ref{eq:p+4he}) without the spin-orbit term when $\rho = 0$.
The obtained values of $\beta_0$, $\beta_1$, and $\beta_W$ are presented in 
Sec.~\ref{sec:results}.

\subsection{Single-folding model for $\alpha$-nucleus system}

The single-folding model has often been applied to construct the $\alpha$-nucleus potential~\cite{BER80, SAT97, KOL00}.
Here we explain the procedure to obtain the $\alpha$-nucleus potential using the present DD-$\alpha N$ interaction.
The $\alpha$-nucleus FMP, $U_{\alpha A}(R, E/A)$, is calculated by
\begin{equation}
U_{\alpha A}(R, E/A) = \int{ U_{{\rm DD-}\alpha N}(s, \rho, E/A)\ \rho (r) \ d\bm{r}}, \label{eq:fold}
\end{equation}
where $\bm{R}=\bm{s}+\bm{r}$.
$\bm{s}$ is the distance vector between the $\alpha$ particle and a nucleon in the target nucleus.
$\rho (r)$ is the point-nucleon density of the target nucleus.
In this paper, we employ the point-nucleon density obtained by the mean-field model, that is, the Skyrme-HF+BCS model~\cite{EBA10} with the SkM* parameter set~\cite{BQB82}.
The center of mass correction is made in the same manner as Ref.~\cite{NEG70}.
These density distributions were already applied to construct the global OMP of the proton-nucleus elastic scattering showing good reproduction of various cross-section data in a wide range of incident energies~\cite{FUR19}.
We evaluate $\rho$ in $U_{{\rm DD-}\alpha N}(s, \rho, E/A)$ as the position of the interacting nucleon in the target nucleus, i.e., $\rho = \rho (\bm{r})$.
The integration in Eq.~(\ref{eq:fold}) can easily be performed by using the Fourier transformation.

\section{Results and Discussion}
\label{sec:results}

In this section, the calculated results are presented.
In the cross-section calculations, we use the standard Coulomb potential with the uniform charge radius,  $R_C = 1.3 \cdot 4^{1/3}$ fm for $p$ + $^{4}$He systems and $R_C = 1.3 \left( 4^{1/3} + A^{1/3} \right)$ fm for $\alpha$ + nucleus systems with $A$ being the mass number of the target nucleus.
Relativistic kinematics is used to compute the cross sections.

We demonstrate that the present DD-$\alpha N$ interaction is helpful to analyze the $\alpha$-nucleus system in a wide range of incident energies and target nuclei.
We first construct the effective $\alpha N$ interaction that reproduces the $p$ + $^4$He elastic scattering.
After that, we fix the density dependence of the $\alpha N$ interaction so as to reproduce the $\alpha$-nucleus elastic scattering cross-section data.
The total reaction cross sections are computed and compared with the experimental data to verify our model.
Then, we show some examples of the applications of the DD-$\alpha N$ interaction to the $\alpha$ elastic scattering in the wide ranges of the target mass and the incident energy.

\subsection{$p$ + $^4$He elastic scattering}

We calculate the $p$ + $^4$He elastic scattering cross section with the OMP in the form of Eq.~(\ref{eq:p+4he}).
The parameters of the potential are fixed to reproduce the $p$ + $^4$He elastic scattering data for each incident energy $(E/A)$ with the Automatic Local Potential Search (ALPS) code~\cite{ALPS}.
The resulting parameters are given in the following:
\begin{eqnarray}
V_0 &=& -2.31 (E/A)^{1/2} + 78.5 \ \ \ {\rm (MeV)}, \label{param_begin}\\
r_0 &=&  0.00153 (E/A) -0.0798 (E/A)^{1/2} + 1.5 \ \ \ {\rm (fm)}, \\
a_0 &=& 0.0269 (E/A)^{1/2} + 0.262 \ \ \ {\rm (fm)}, \\
V_1 &=& -3.71 (E/A)^{1/2} +14.6 \ \ \ {\rm (MeV)}, \\
r_1 &=& 0.8 \ \ \ {\rm (fm)}, \\
a_1 &=& 0.0044 (E/A)^{1/2} + 0.19 \ \ \ {\rm (fm)}, \\
W &=& 1.6 (E/A)^{1/2} - 0.51 \ \ \ {\rm (MeV)}, \\
r_w &=& -0.0116 (E/A)^{1/2} + 1.18 \ \ \ {\rm (fm)}, \\
a_w &=& 0.5 \ \ \ {\rm (fm)}, \\
V_{\rm SO} &=& -0.423 (E/A)^{1/2} + 12 \ \ \ {\rm (MeV)}, \\
r_{\rm SO} &=& 0.7 \ \ \ {\rm (fm)}, \\
a_{\rm SO} &=& 0.3 \ \ \ {\rm (fm)}. \label{param_end}
\end{eqnarray}

\begin{figure}[ht]
\centering
\includegraphics[width=6.4cm]{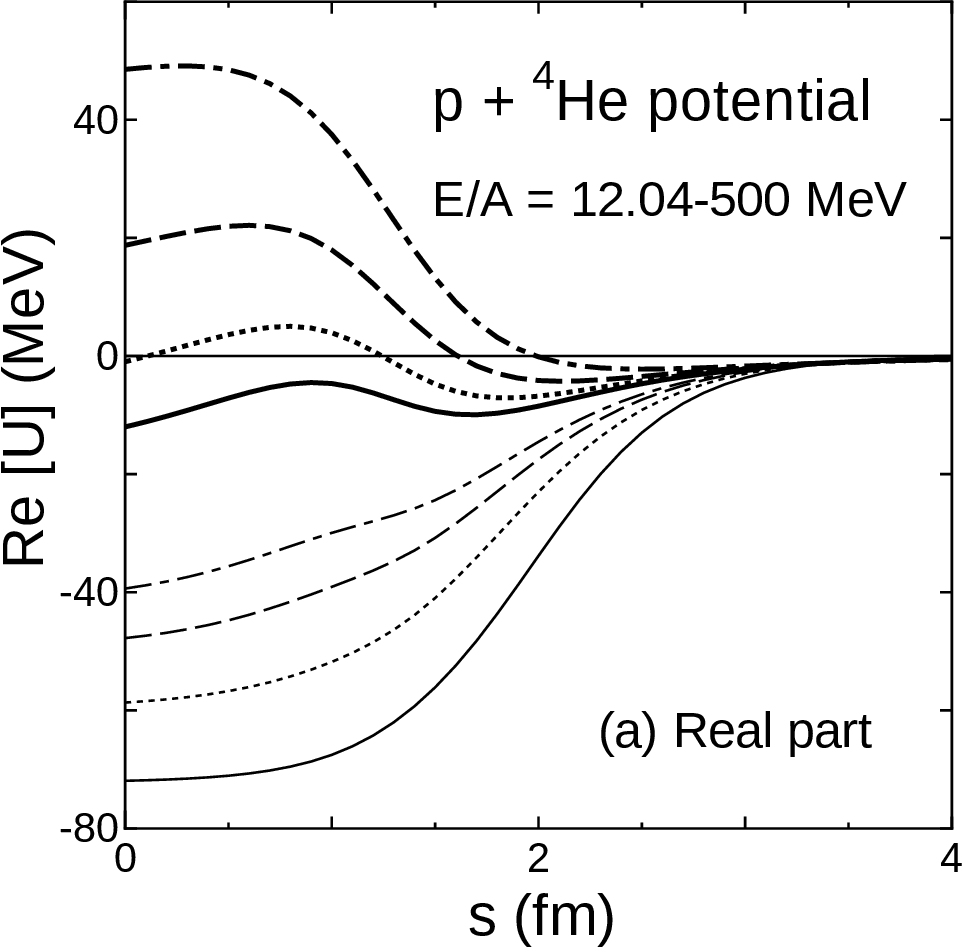}
\hspace{10mm}
\includegraphics[width=6.4cm]{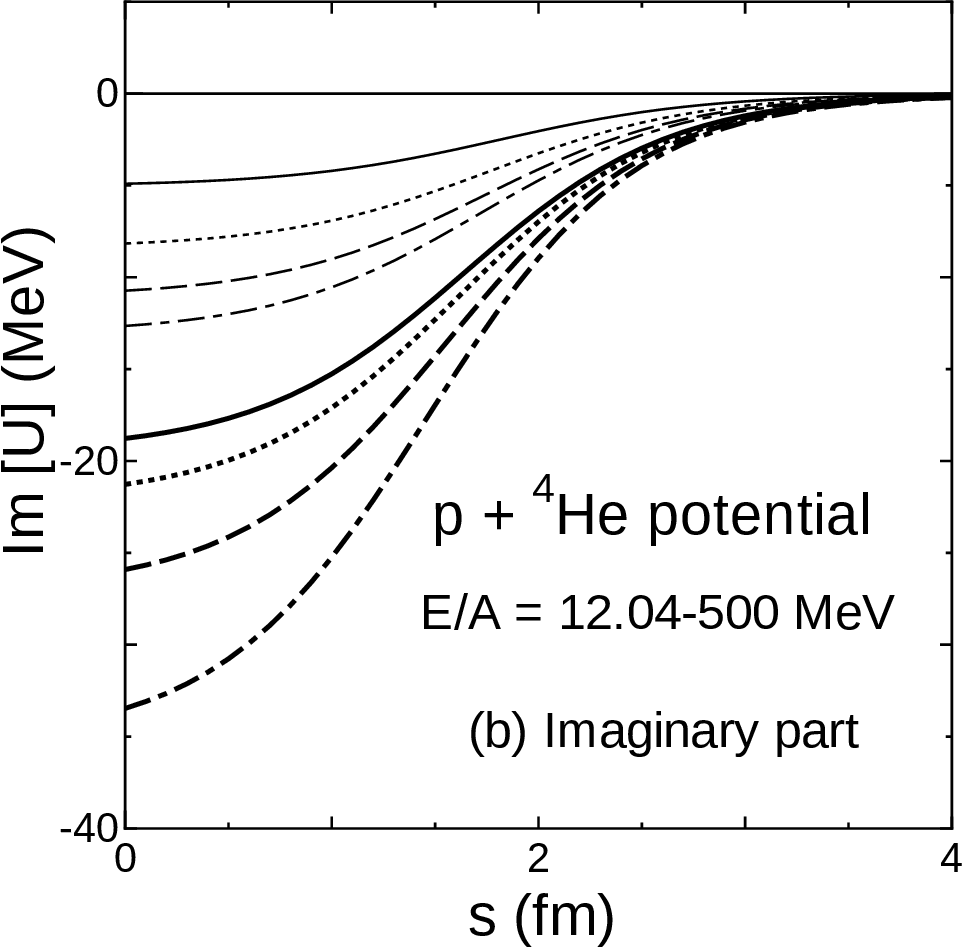}
\caption{(a) Real and (b) imaginary part of OMPs for the $p$ + $^4$He system at $E/A =$ 12.04--500 MeV.
See text for details.}
\label{fig:pot-p+4he}
\end{figure}
Figure~\ref{fig:pot-p+4he} shows the (a) real and (b) imaginary parts of the central terms of the OMP of the $p$ + $^4$He system at 12.04--500 MeV.
The solid, dotted, dashed, dot-dashed, thick-solid, thick-dotted, thick-dashed, and thick-dot-dashed curves indicate the OMPs at $E/A$ = 12.04, 31, 52.3, 71.9, 156, 200, 297, and 500 MeV, respectively.
Both the real and imaginary parts show smooth energy dependence.
It should be noted that the complicated energy dependence of the OMP, i.e., the wine-bottle shape to repulsive potential, is well described by introducing the double WS type potential for the real part.
It is expected that this repulsive potential has an important role in describing not only the $p$ + $^4$He elastic scattering but also the $\alpha$-nucleus elastic scattering as discussed in Refs.~\cite{HAM90, SEN87, FUR10}.
For the imaginary part, the energy dependence of the potential is relatively mild.
Note that in this paper the surface term, which is usually included in heavier targets, is not introduced for the imaginary potential.
We confirmed that the term gives only a minor contribution, which does not affect the fitting results and can be omitted.

\begin{figure}[ht]
\centering
\includegraphics[width=6.4cm]{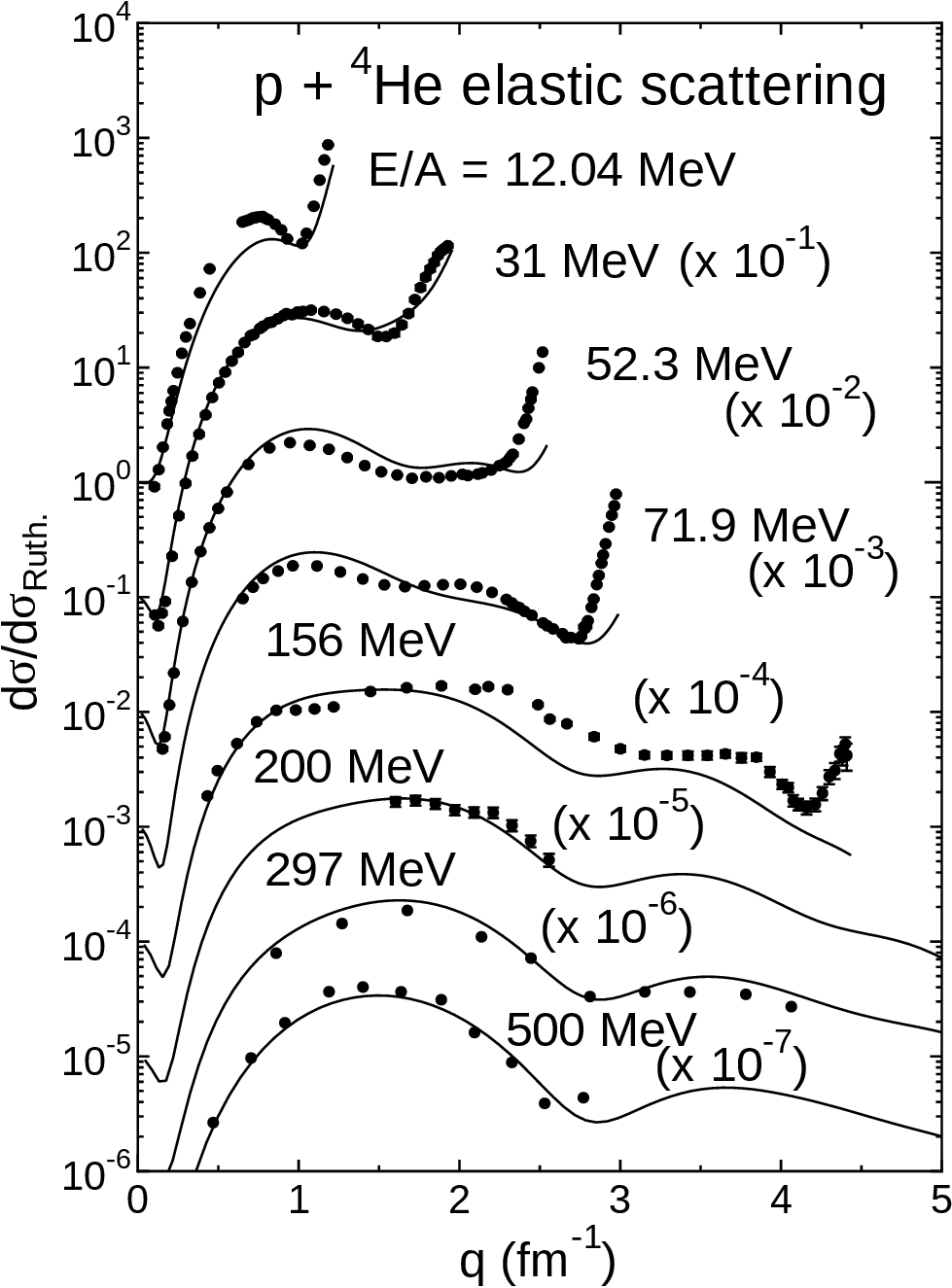}
\caption{Rutherford ratio of the elastic scattering cross sections of the $p$ + $^4$He system at $E/A =$ 12.04--500 MeV as a function of the momentum transfer $q$.
The solid curves are the results obtained by the present work.
The experimental data are taken from~\cite{EXFOR, SAN59, BUN64, IMA79, BUR89, COM75, CHE18, YOS01, STE92}.}
\label{fig:p+4he}
\end{figure}

Figure~\ref{fig:p+4he} compares the calculated and experimental elastic scattering cross sections of the $p$ + $^4$He system at $E/A =$ 12.04--500 MeV.
The solid curves are the results obtained by the present work.
We see that with the parametrization given in Eqs.~(\ref{param_begin})--(\ref{param_end}), the $p$ + $^{4}$He elastic scattering cross sections are well reproduced.
At high incident energies, the repulsive core introduced in Eq.~(\ref{eq:p+4he}) is needed to reproduce the cross-section data.
Though there are some discrepancies between the experimental data and the calculated results, the central part of the $p$ + $^4$He OMP is reasonably fixed because the discrepancies can be improved by modifying the spin-orbit part for each incident energy, which does not affect the $\alpha$-nucleus OMP.

\subsection{Determination of density dependence by $\alpha$-nucleus elastic scattering}

In this subsection, we fix the density-dependent part of the present DD-$\alpha N$ interaction by the $\alpha$-nucleus elastic scattering.
There are plenty of experimental data for the $\alpha$-nucleus elastic scattering.
In this paper, we choose the $^{16}$O, $^{40}$Ca, $^{58}$Ni, $^{90}$Zr, and $^{208}$Pb nuclei as a target nucleus because their reactions are simple as their excitation energies are large and the channel-coupling effect is expected to be small.

We obtain the $\alpha$-nucleus potential by the folding procedure given by Eq.~(\ref{eq:fold}).
Though we construct the potential based on the $p$ + $^4$He reaction, the imaginary part of the $\alpha$-nucleus OMP should be modified despite that the density dependence is introduced because the imaginary part of the $\alpha$-nucleus potential also includes the contribution from the target nucleus.
Therefore, we apply the renormalization factor $N_I$ to the imaginary part as 
\begin{eqnarray}
U_{\alpha A}(R; E/A) &=& V_{\alpha A}(R; E/A) + i W_{\alpha A}(R; E/A) \\
&\to& V_{\alpha A}(R; E/A) + i N_I W_{\alpha A}(R; E/A),
\end{eqnarray}
where $V_{\alpha A}(R; E/A)$ and $W_{\alpha A}(R; E/A)$ are the real and imaginary parts of the constructed $\alpha$-nucleus potential with the present DD-$\alpha N$ interaction.

\begin{figure}[ht]
\centering
\includegraphics[width=6.4cm]{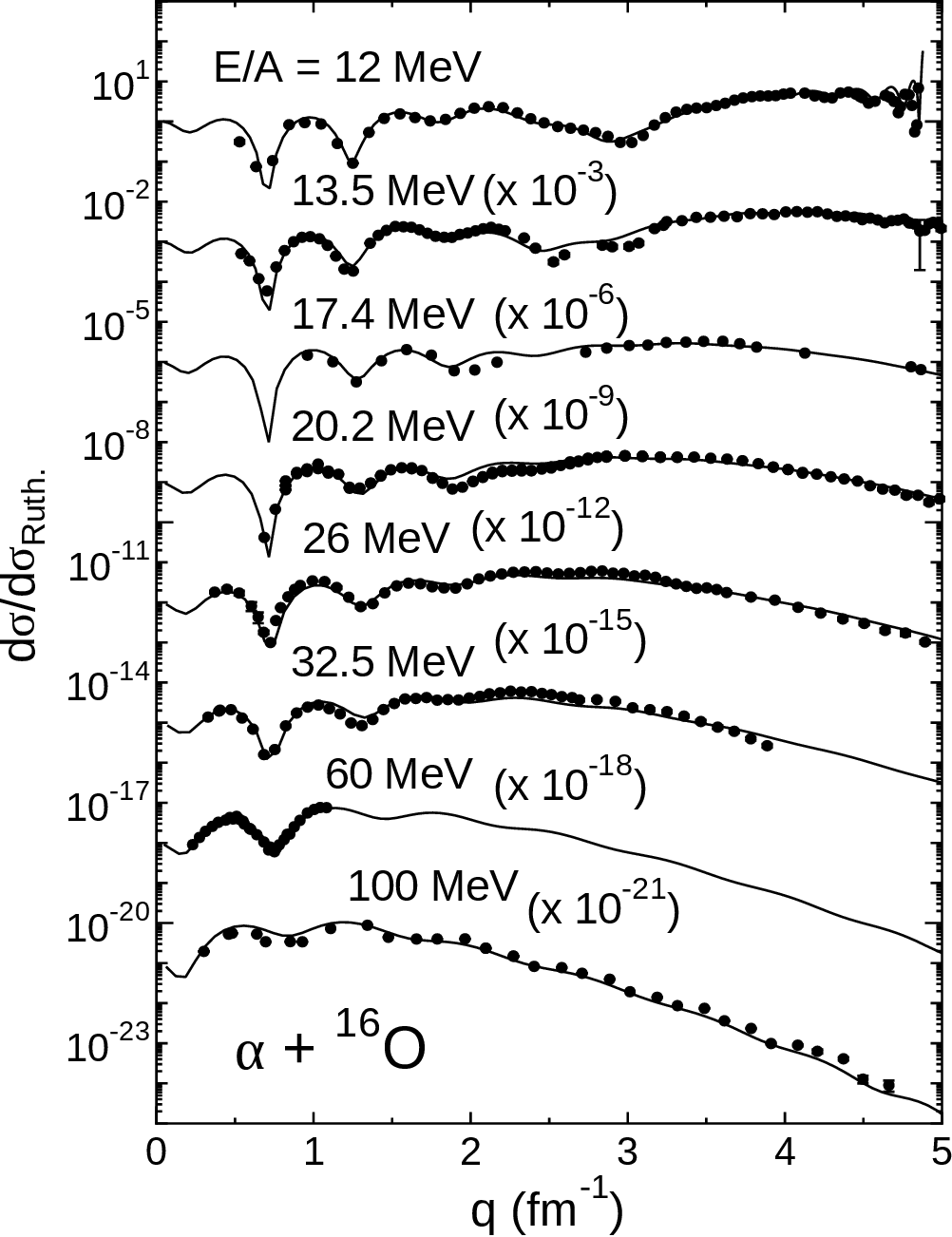}
\caption{Elastic scattering cross sections of $\alpha + {}^{16}$O system at $E/A$ = 12--100 MeV.
The solid curves are the results obtained by the present calculation.
The experimental data are taken from \cite{EXFOR, BUR17, ABE87, MIC83, HAU69, ADA18, LUI01, WAK07}.}
\label{fig:o-target}
\end{figure}

\begin{figure}[ht]
\centering
\includegraphics[width=6.4cm]{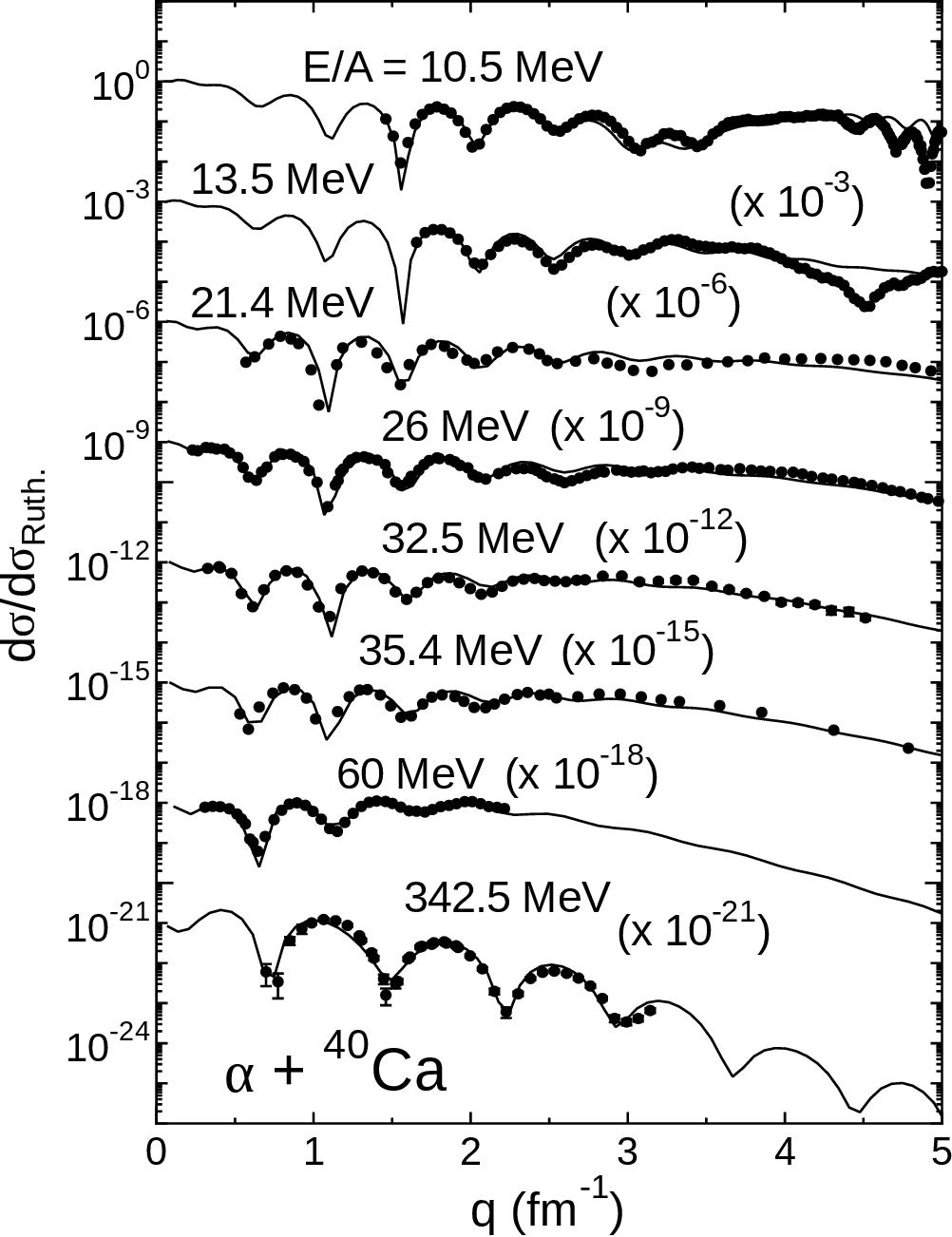}
\caption{Elastic scattering cross sections of $\alpha + {}^{40}$Ca system at $E/A$ = 10.5--342.5 MeV.
The solid curves are the results obtained by the present calculation.
The experimental data are taken from \cite{EXFOR, GUB81, CHA76, GIL80, ADA18, GOL74, YOU97, ALK77}.}
\label{fig:ca-target}
\end{figure}

\begin{figure*}[htb]
\centering
\includegraphics[width=5.4cm]{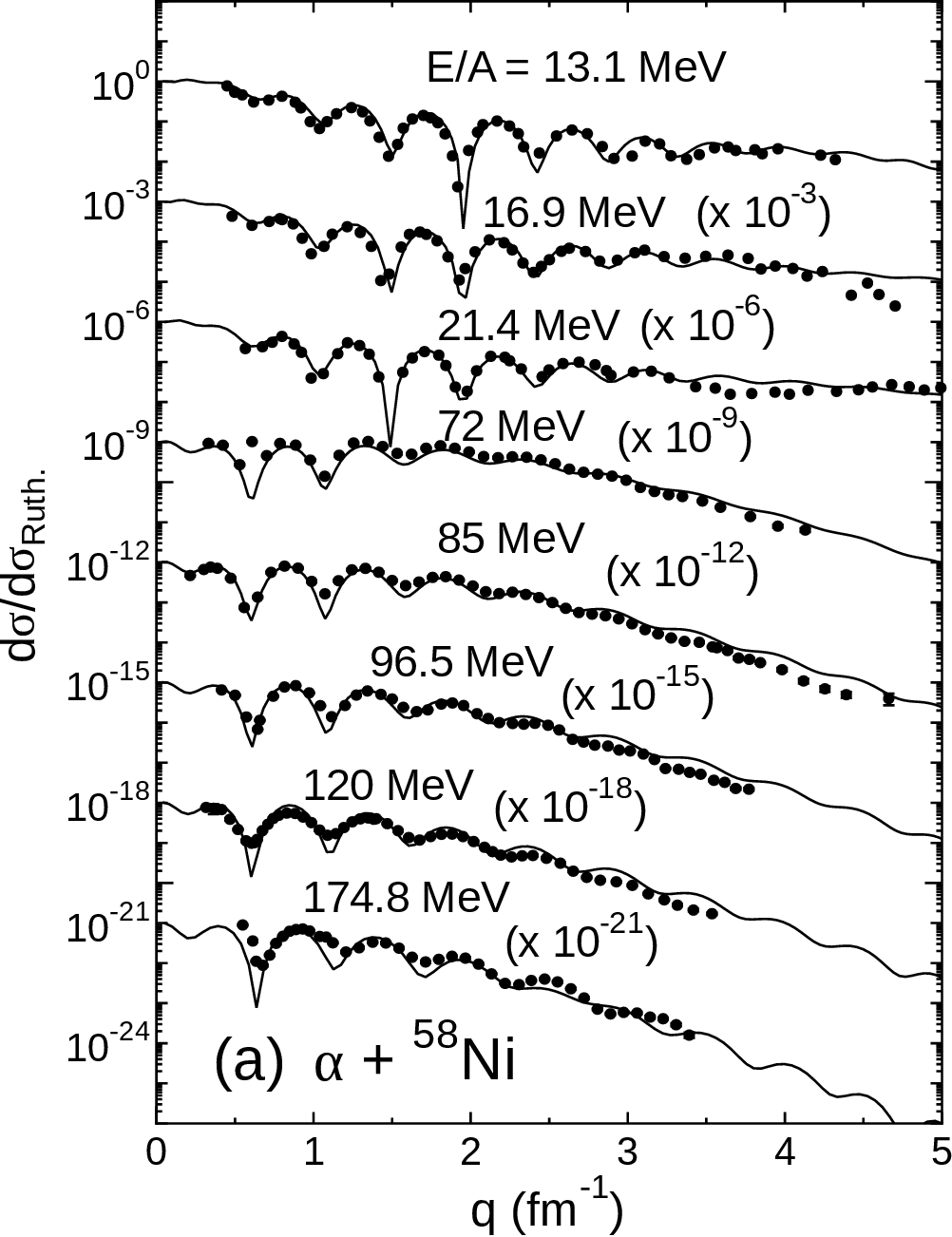}
\includegraphics[width=5.4cm]{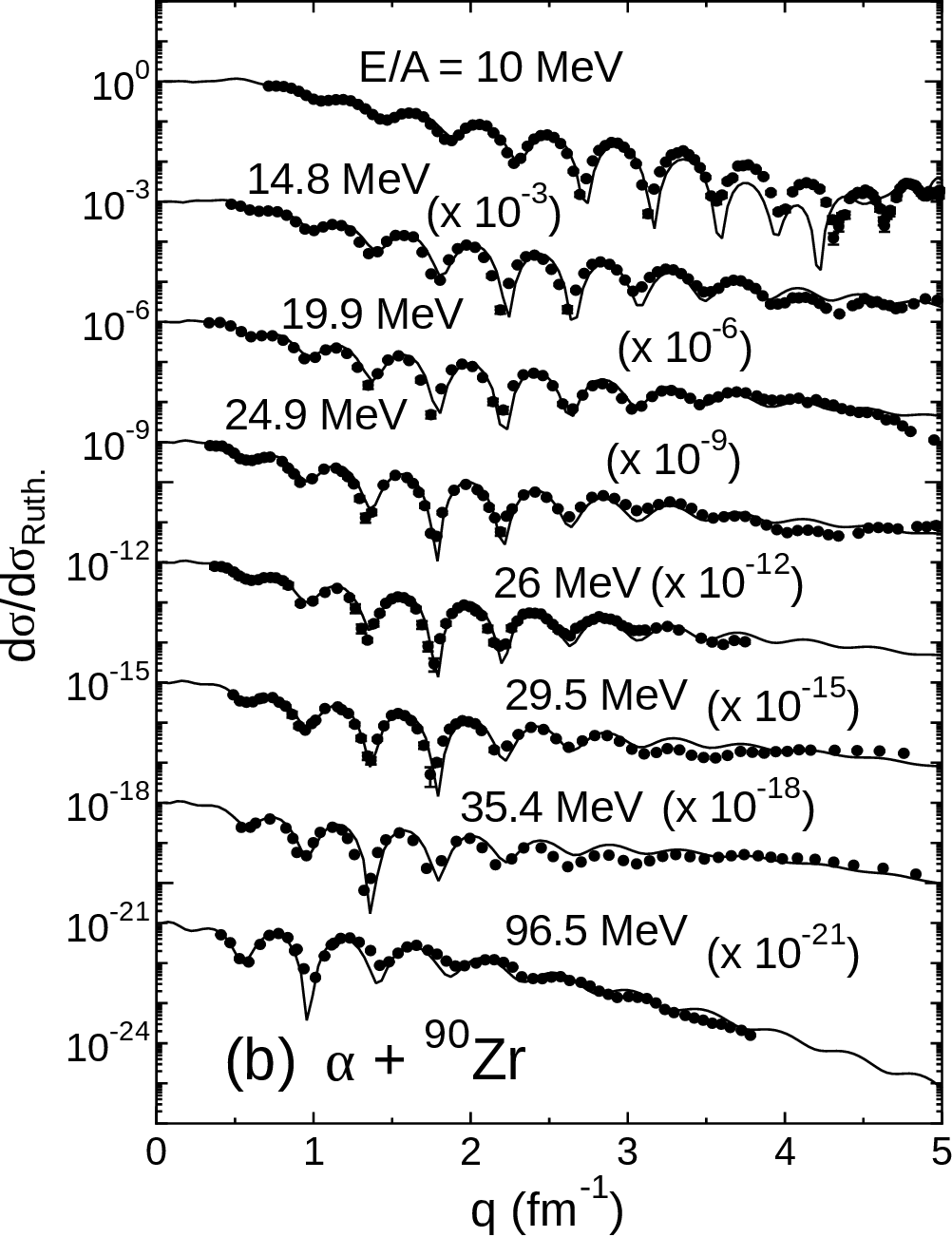}
\includegraphics[width=5.4cm]{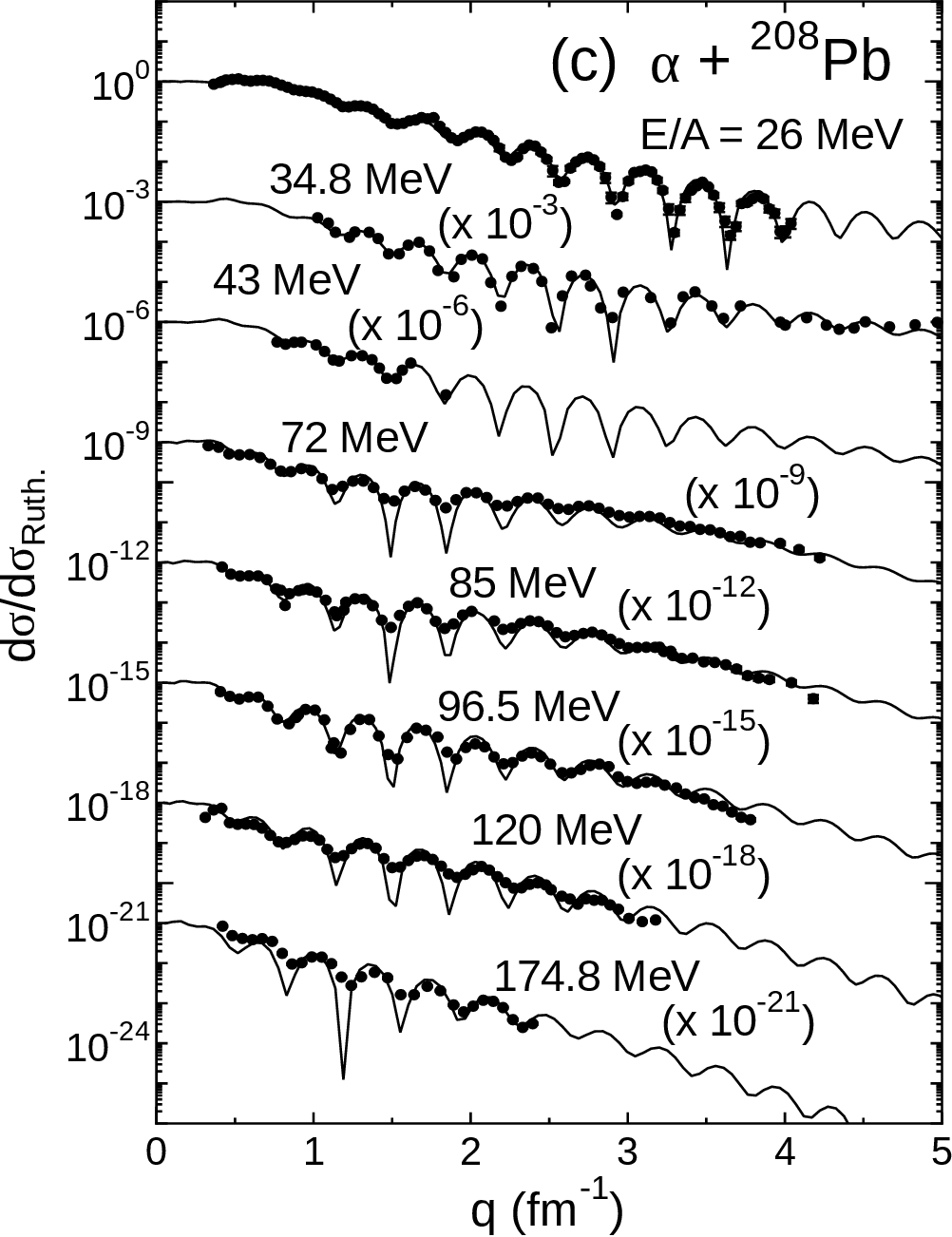}
\caption{Elastic scattering cross sections of (a) $\alpha + {}^{58}$Ni system at $E/A$ = 13.1--174.8 MeV, (b) $\alpha + {}^{90}$Zr system at $E/A$ = 10--96.5 MeV, and (c) $\alpha + {}^{208}$Pb system at $E/A$ = 26--174.8 MeV.
The solid curves are the results obtained by the present calculation.
The experimental data are taken from \cite{EXFOR, CHA76, BON85, UCH04, PUT77, HAU69, GOL74, GOL73, MOR80}.}
\label{fig:hi-target}
\end{figure*}

We fix the parameters introduced in Eqs.~(\ref{eq:bd-real1}), (\ref{eq:bd-real2}), and (\ref{eq:bd-imag}) to reproduce the experimental data as shown in Figs.~\ref{fig:o-target}, \ref{fig:ca-target}, and \ref{fig:hi-target}.
Finally, we obtain
\begin{eqnarray}
\beta_0 (E/A) &=& 6.56 \exp{\left( -\frac{(E/A)^2}{65^2} \right)} + 2.3 \ \ \ {\rm (fm^3)}, \label{eq:beta0}\\
\beta_1 &=& 4.3 \ \ \ {\rm (fm^3)}, \label{eq:beta1}\\
\beta_W &=& 9.8 \ \ \ {\rm (fm^3)}. \label{eq:betaW}
\end{eqnarray}
Note that the energy dependence for $\beta_0$ is introduced to obtain better reproduction for the data, implying complicated medium effects in the $\alpha$-nucleus scattering, which is difficult to derive theoretically at this moment.
Here we discuss the $\beta_0$ and $\beta_1$ values in detail.
The $\beta_0$ value rapidly decreases as increasing the incident energy in the ranges of $10 < E/A < 100$ MeV.
In the energy region, the effect of the core part of the $p$ + $^4$He potential and its density dependence ($\beta_1$) is negligible.
As the energy increases, though the $\beta_0$ value is almost constant, the core part of the $p$ + $^4$He potential and its density dependence ($\beta_1$) become important.

If the small $\beta_0$ value ($\beta_0$ = 2--3) is applied for the lower energy, the FMP becomes deep, which results in the overestimation at the backward angles.
This overestimation cannot be corrected by adjusting other parameters.
On the other hand, the elastic scattering cross sections cannot be reproduced 
when the large $\beta_0$ value ($\beta_0$ = 7--8) is applied to $E/A =$ 100 MeV.
Therefore, we decided to give the energy dependence on the $\beta_0$ value (a medium effect).
Note that a few experimental data for higher energy regions are available, only at $E/A =$ 342.5 MeV.
The density dependence ($\beta_1$) can be refined if experimental cross-section data for the high incident energies are provided.

In the following, we compare the calculated and experimental cross sections for the selected nuclei.
Figure~\ref{fig:o-target} shows the $\alpha$ + $^{16}$O elastic scattering cross sections at $E/A$ = 12--100 MeV as a function of the momentum transfer $q=2k\sin(\theta/2)$ with the wave number $k$ and center-of-mass scattering angle $\theta$.
The calculated cross sections agree well with the experimental data up to the backward angles.
We can obtain a useful $\alpha$-nucleus potential with the present DD-$\alpha N$ interaction if $N_I$ is properly chosen.
\begin{figure}[ht]
\centering
\includegraphics[width=6.4cm]{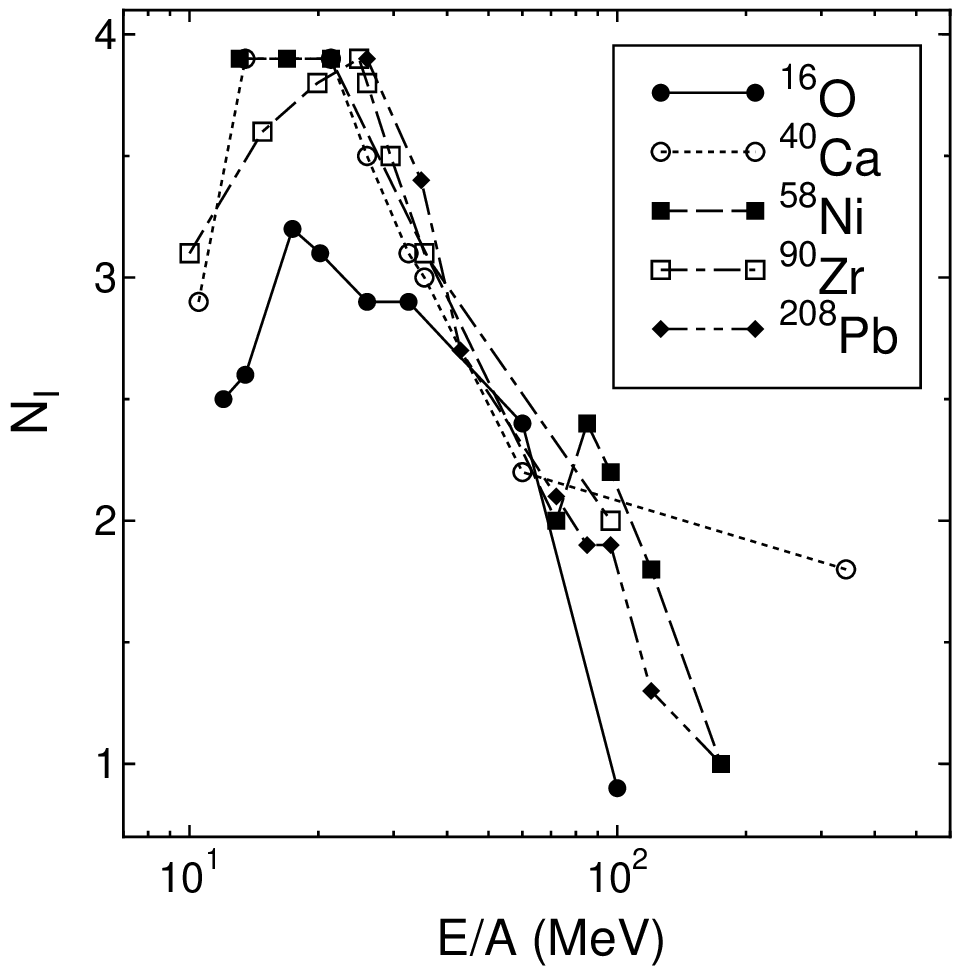}
\caption{$N_I$ values adopted for the $\alpha$ + $^{16}$O, $^{40}$Ca, $^{58}$Ni, $^{90}$Zr, and $^{208}$Pb systems.}
\label{fig:NI}
\end{figure}
The $N_I$ values for the $\alpha$ + $^{16}$O system are described in Fig.~\ref{fig:NI}.
We see the energy dependence also for the $N_I$ value.
We will discuss it later after the results for heavier target nuclei are presented and in Sec.~\ref{sec:global}.

Figure~\ref{fig:ca-target} shows the results of the $\alpha$ + $^{40}$Ca elastic scattering at $E/A$ = 10.5--342.5 MeV.
As we already see for the $\alpha$ + $^{16}$O scattering, the agreement between the theory and experiment is satisfactory.
At 13.5 MeV, we see a slight discrepancy between the theoretical result and the experimental data at backward angles although the $\alpha$ + $^{16}$O elastic scattering cross sections are reproduced up to backward angles at low incident energies.
We will discuss this matter in more detail later.
At $E/A =$ 342.5 MeV, we find that the core part of the $p$ + $^4$He potential and its density dependence ($\beta_1$) are important to reproduce the experimental data.
The different patterns of the angular distributions are clearly seen for $E/A <$ 60 MeV and $E/A =$ 342.5 MeV. 
At $E/A =$ 342.5 MeV, the strong interference by the nearside and farside ($N/F$) components is clearly seen. 
In such a situation, the real part of the OMP is no longer a normal attractive potential.
The potential shape changes from attractive to repulsive as increasing the incident energy.
When this transition occurs, the characteristic interference of the $N/F$ components is seen in the angular distribution.
The relationship between the repulsive nature of the potential and the elastic scattering cross section was discussed in detail in Ref.~\cite{FUR10}.
Here, there are a few experimental data at a higher incident energy ($E/A >$ 200 MeV) for the $\alpha$ elastic scattering.
Such experimental data is needed for a more detailed analysis.

Figure~\ref{fig:hi-target} compares the theoretical and experimental cross sections of (a) the $\alpha$ + $^{58}$Ni elastic scattering at $E/A$ = 13.1--174.8 MeV, (b) the $\alpha$ + $^{90}$Zr elastic scattering at $E/A$ = 10--96.5 MeV, and (c) the $\alpha$ + $^{208}$Pb elastic scattering at $E/A$ = 26--174.8 MeV, respectively.
We also see the validity of the present DD-$\alpha N$ interaction.
The density dependence is adequately represented for both the lower and higher incident energies.

Here, we discuss the $N_I$ value, which is introduced to readjust the strength of the imaginary potential.
As we see the comparison in the various $\alpha$-nucleus elastic scattering cross-section data, the potential form of the imaginary potential is reasonably obtained by the present DD-$\alpha N$ interaction.
The $N_I$ values for $\alpha$ + $^{16}$O system as 2.8, 2.8, 3.5, 3.1, 2.9, 2.9, 2.0, and 0.8 at $E/A$ = 12, 13.5, 17.4, 20.2, 26, 32.5, 60, and, 100, respectively.
For the $^{40}$Ca target, the $N_I$ values are 2.8, 4.3, 4.2, 3.7, 3.2, 3.0, 2.2, and 1.9 at $E/A$ = 10.5, 13.5, 21.4, 26, 32.5, 35.4, 60, and 342.5, respectively.
For the $^{58}$Ni target, the $N_I$ values are 3.9, 4.3, 4.3, 2.0, 2.4, 2.3, 2.2, and 1.1 at $E/A$ = 13.1, 16.9, 21.4, 72, 85, 96.5, 120, and 174.8, respectively.
For the $^{90}$Zr target, the $N_I$ values are 2.8, 4.0, 4.2, 4.2, 4.2, 3.6, 3.0, and 2.4 at $E/A$ = 10, 14.8, 19.9, 24.9, 26, 29.5, 35.4, and 96.5, respectively.
For the $^{208}$Pb target, the $N_I$ values are 4.6, 3.8, 3.7, 2.3, 2.2, 2.3, 2.1, and 1.0 at $E/A$ = 26, 34.8, 43, 72, 85, 96.5, 120, and 174.8, respectively.
As shown in Fig.~\ref{fig:NI}, the energy dependence of the $N_I$ values is similar in all the systems.
The $N_I$ value becomes smaller as the energy increases and becomes larger as the mass number increases.
This behavior may be attributed to the differences in the level densities as the nuclear system approaches nuclear matter at high incident energy and large mass number. 

\subsection{Total reaction cross sections}
\begin{figure}[ht]
\centering
\includegraphics[width=6.4cm]{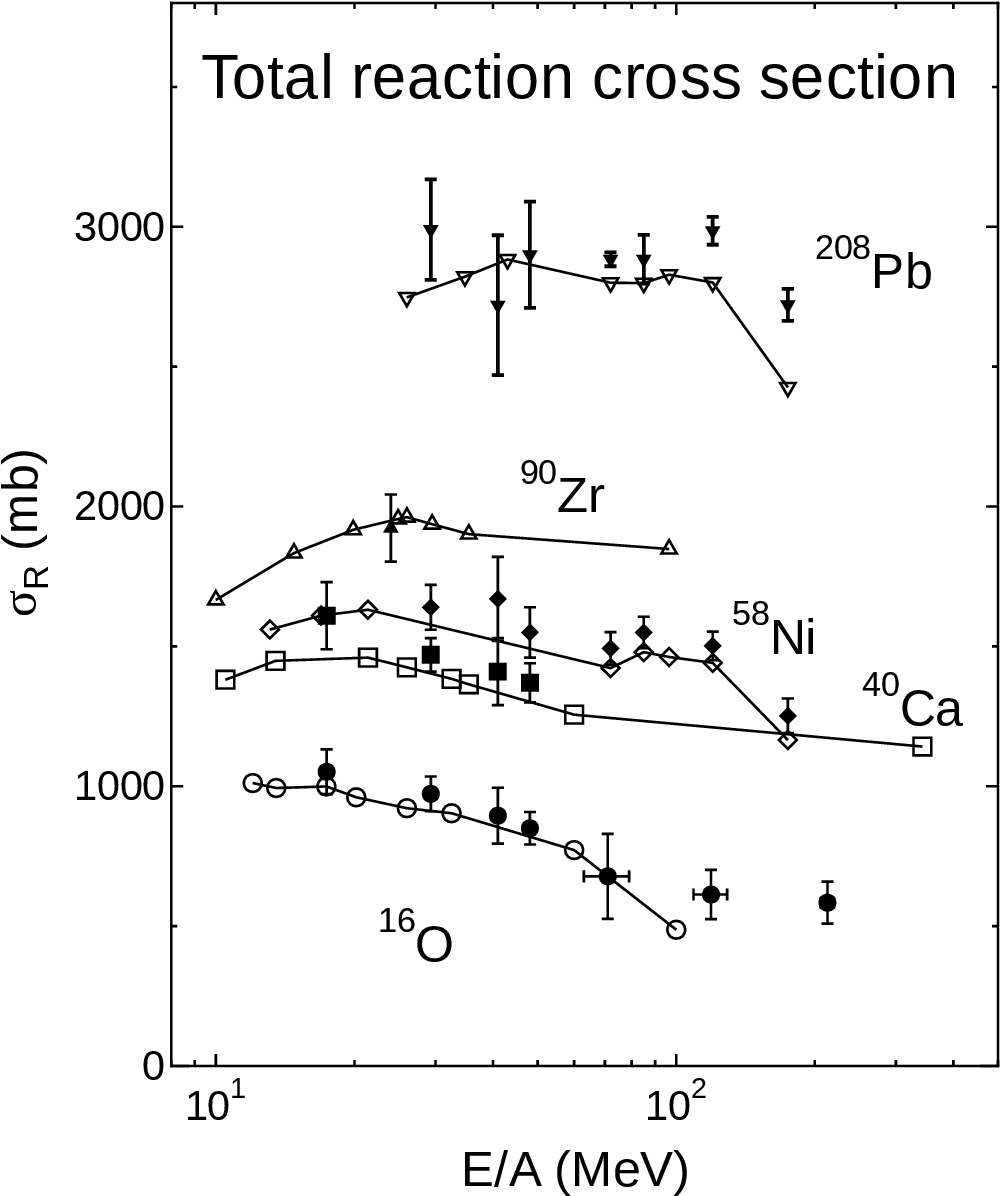}
\caption{Total reaction cross sections for $\alpha$-nucleus scattering.
The circles, squares, diamonds, triangles, and inverted triangles are the results of the $^{16}$O, $^{40}$Ca, $^{58}$Ni, $^{90}$Zr, and $^{208}$Pb targets, respectively.
The filled symbols with the error bar are the experimental data.
The open symbols are obtained by the present calculation.
The connecting solid lines are a guide to the eyes.
The experimental data are taken from Refs.~\cite{ING00, HOR19, BON85, DUY03}.}
\label{fig:rxs}
\end{figure}
Here, we show the total reaction cross section.
Figure~\ref{fig:rxs} compares the results of the total reaction cross section for the incident $\alpha$ particle for the $^{16}$O, $^{40}$Ca, $^{58}$Ni, $^{90}$Zr, and $^{208}$Pb targets.
The circles, squares, diamonds, triangles, and inverted triangles are the results of the $^{16}$O, $^{40}$Ca, $^{58}$Ni, $^{90}$Zr, and $^{208}$Pb targets, respectively.
The filled symbols with the error bar are the experimental data.
The open symbols are the result obtained by the present DD-$\alpha N$ interaction.
The solid lines connecting to the calculated cross sections are drawn for a guide to the eyes.
Almost all cross sections are reproduced by the present $\alpha$-nucleus potential.
We notice the slight underestimation of the $^{208}$Pb cross section at high energy although the elastic scattering cross section is reproduced up to backward angles.
This result indicates that we need care to apply the present DD-$\alpha N$ interaction to the heavier target at high energy.
To understand the nucleus-nucleus interaction in a wide range of the incident energy and target mass from the (semi-)microscopic viewpoint, we need not only the observation of the high-energy nucleus scattering but also the observation of the total reaction cross section at the high-energy heavier target.
Note that the density-dependent parameters and the renormalization factor for the imaginary part are fixed to reproduce the $\alpha$-nucleus elastic scattering cross-section data.
As the total reaction cross sections are reproduced with the same parametrization, the present DD-$\alpha N$ interaction is considered to be global.
Namely, one can apply the present DD-$\alpha N$ interaction to $\alpha$-nucleus reactions only with one free parameter $N_I$. 
The $N_I$ value can be fixed if the total reaction cross-section data is available.

\subsection{Global description of $\alpha$ elastic scattering with present DD-$\alpha N$ interaction}
\label{sec:global}
\begin{figure}[ht]
\centering
\includegraphics[width=5cm]{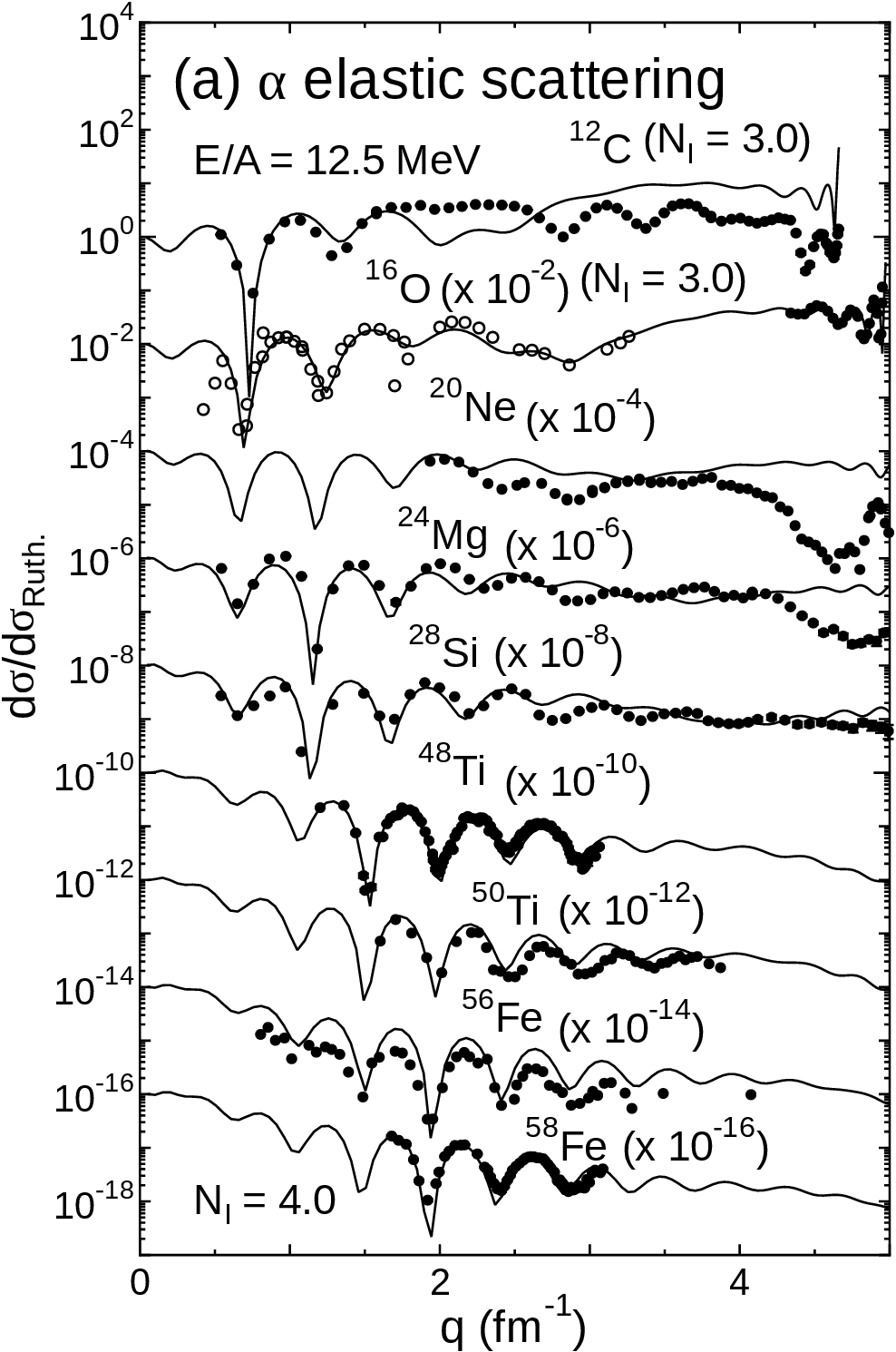}
\includegraphics[width=5cm]{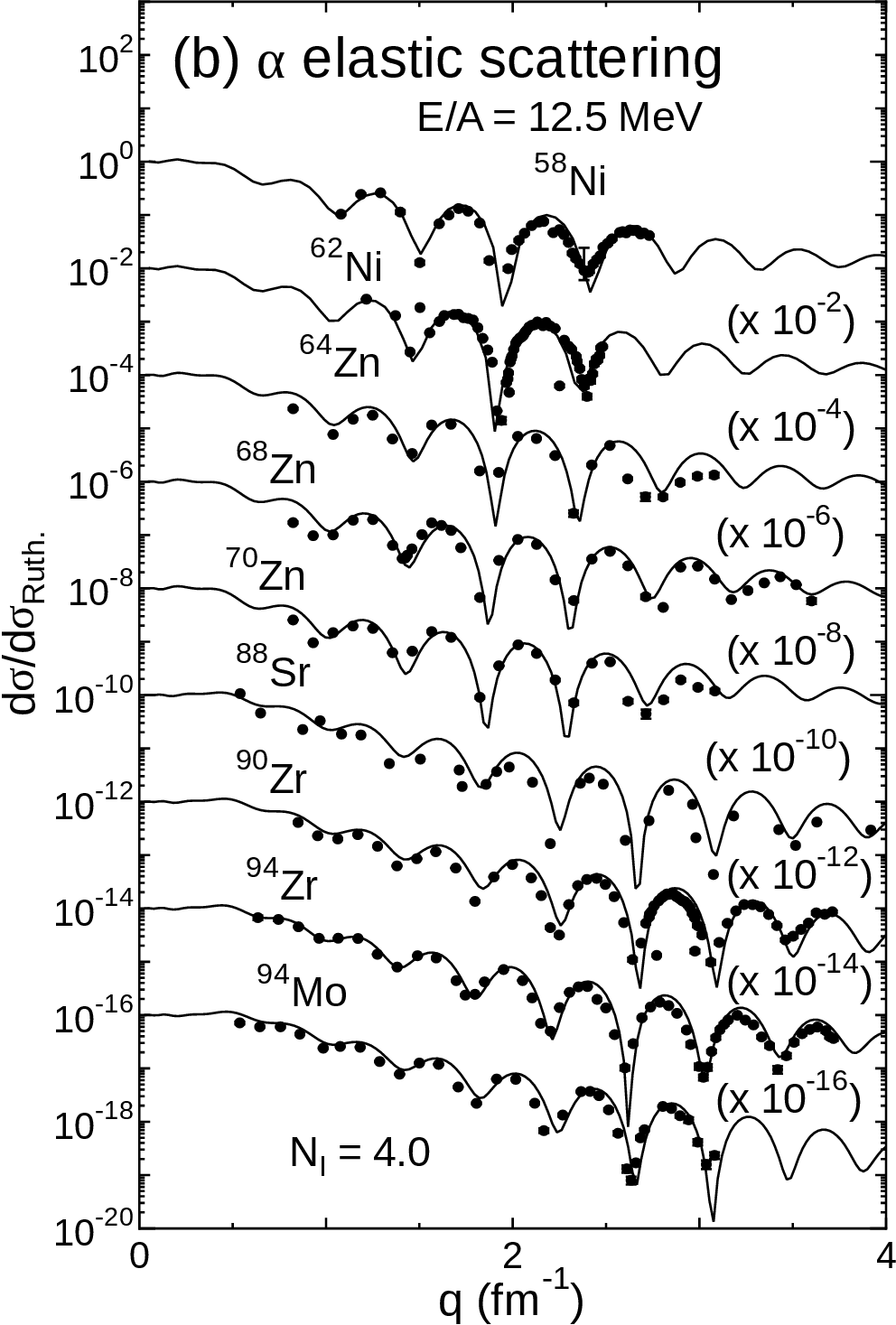}
\includegraphics[width=5cm]{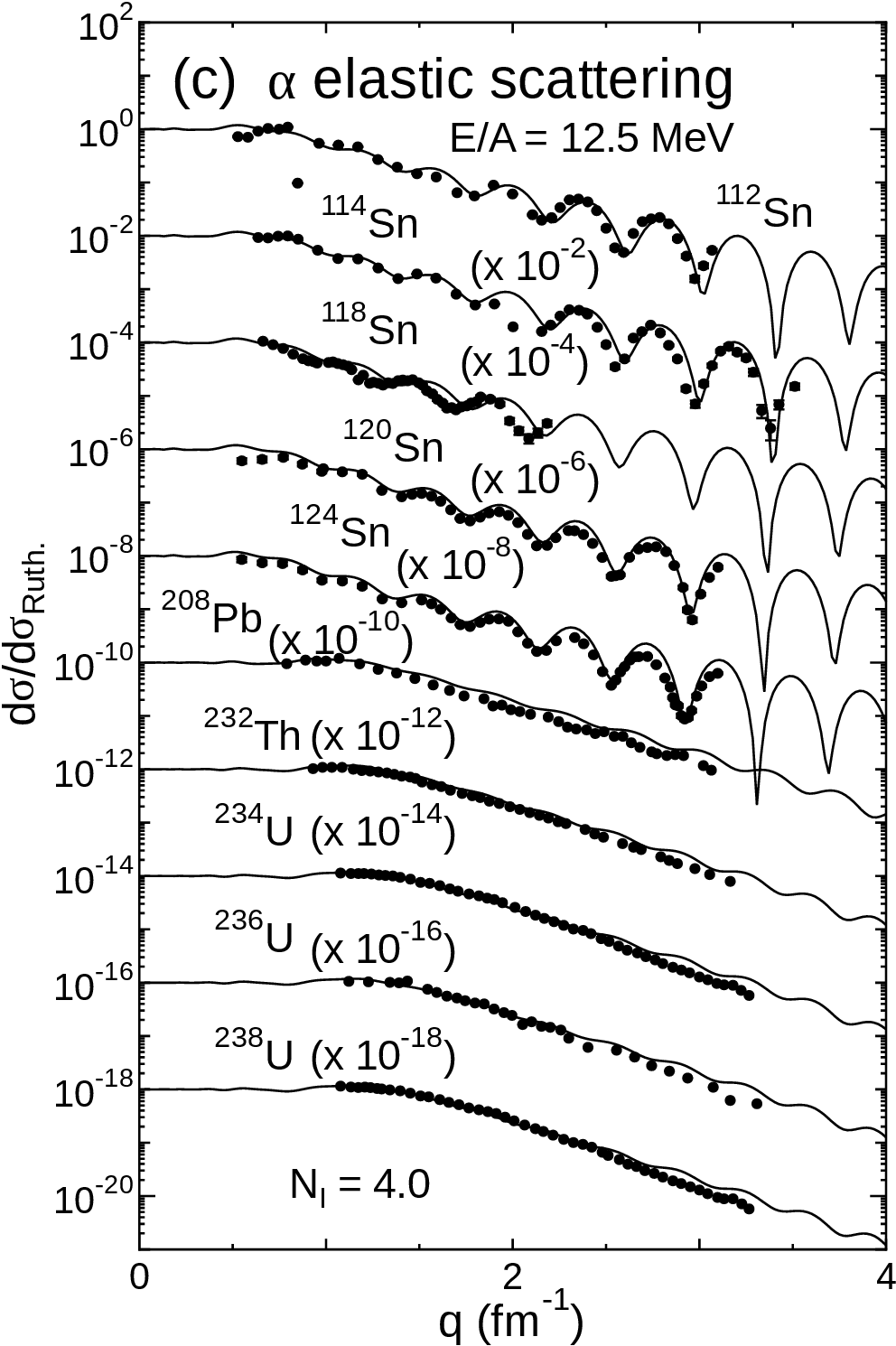}
\caption{Elastic scattering cross sections for $\alpha$-nucleus scattering at $E/A$ = 12.5 MeV.
The experimental data are taken from Refs.~\cite{MIC83,SAD21,DAT89,DUY03,DAV76}.}
\label{fig:xs-low}
\end{figure}
The present DD-$\alpha N$ interaction is constructed based on the $p$ + $^{4}$He elastic scattering and the $\alpha$ + $^{16}$O, $^{40}$Ca, $^{58}$Ni, $^{90}$Zr, and $^{208}$Pb systems.
We here demonstrate the validity of the present DD-$\alpha N$ interaction for $\alpha$-nucleus elastic scattering.
In addition, we discuss how we fix one free parameter ($N_I$).
Figure~\ref{fig:xs-low} shows the $\alpha$ elastic scattering cross sections of $A$ = 12--238 nuclei at $E/A$ = 12.5 MeV.
The $N_I$ values are fixed to be 4.0, which is roughly estimated from Fig.~\ref{fig:NI}.
For $^{12}$C and $^{16}$O targets, the $N_I$ values are fixed to be 3.0 to reproduce the $\alpha$ + $^{16}$O data.
The experimental data are globally reproduced up to backward angles.
However, we see the discrepancy in the results of $^{12}$C, $^{20}$Ne, $^{24}$Mg, and $^{28}$Si targets.
The discrepancy seems to be related to the result of the $\alpha$ + $^{40}$Ca system at 13.5 MeV.
This result implies that we need care when the present DD-$\alpha N$ interaction is applied to the $\alpha$-nucleus elastic scattering at low energies $E/A \sim$ 13 MeV and for $A \leq$ 40 targets though the $\alpha$ + $^{16}$O system is well reproduced.
On the other hand, we find a minor dependence on the $N_I$ value for the target nucleus ($A >$ 40) at $E/A$ = 12.5 MeV.

\begin{figure}[ht]
\centering
\includegraphics[width=5cm]{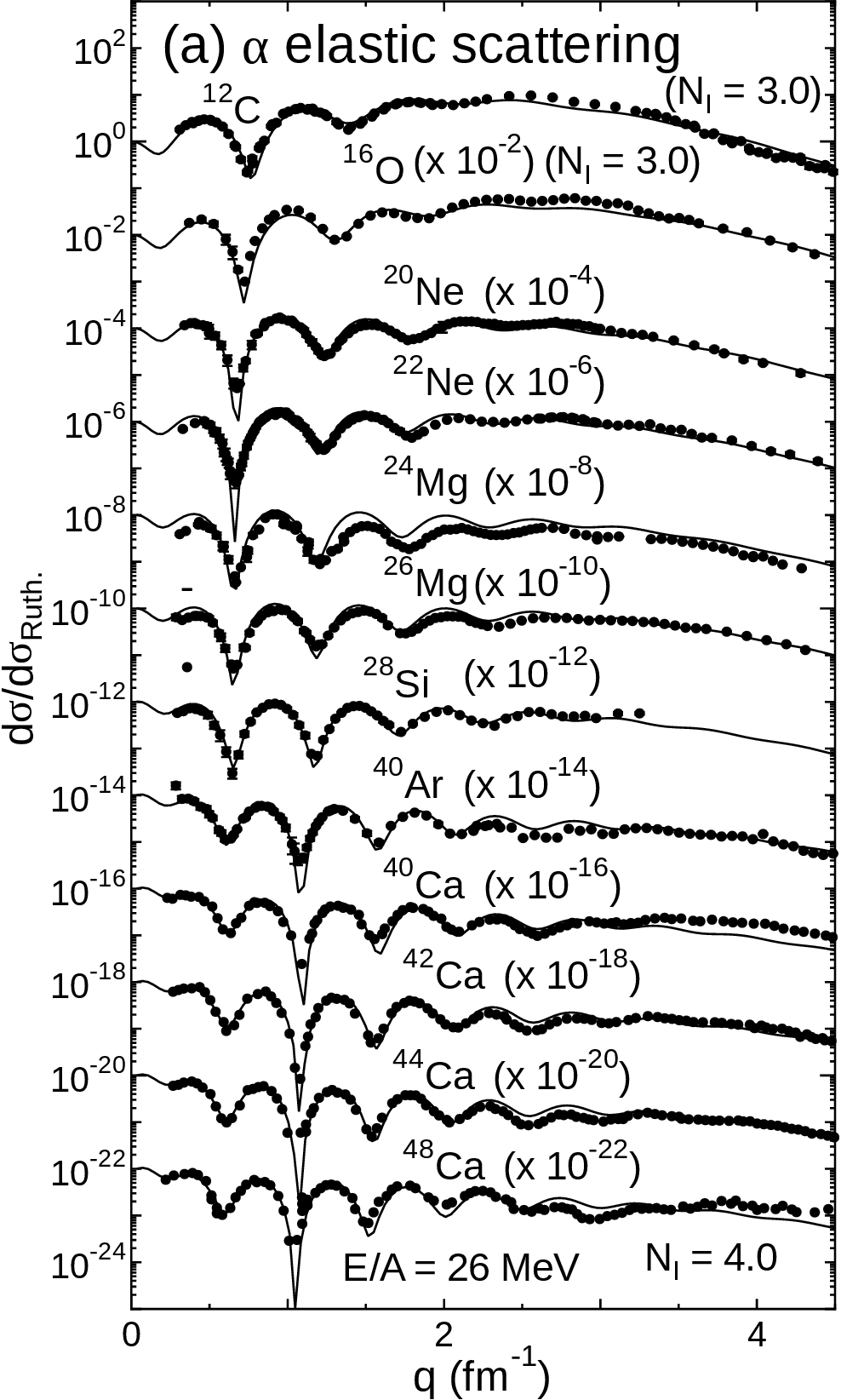}
\includegraphics[width=5cm]{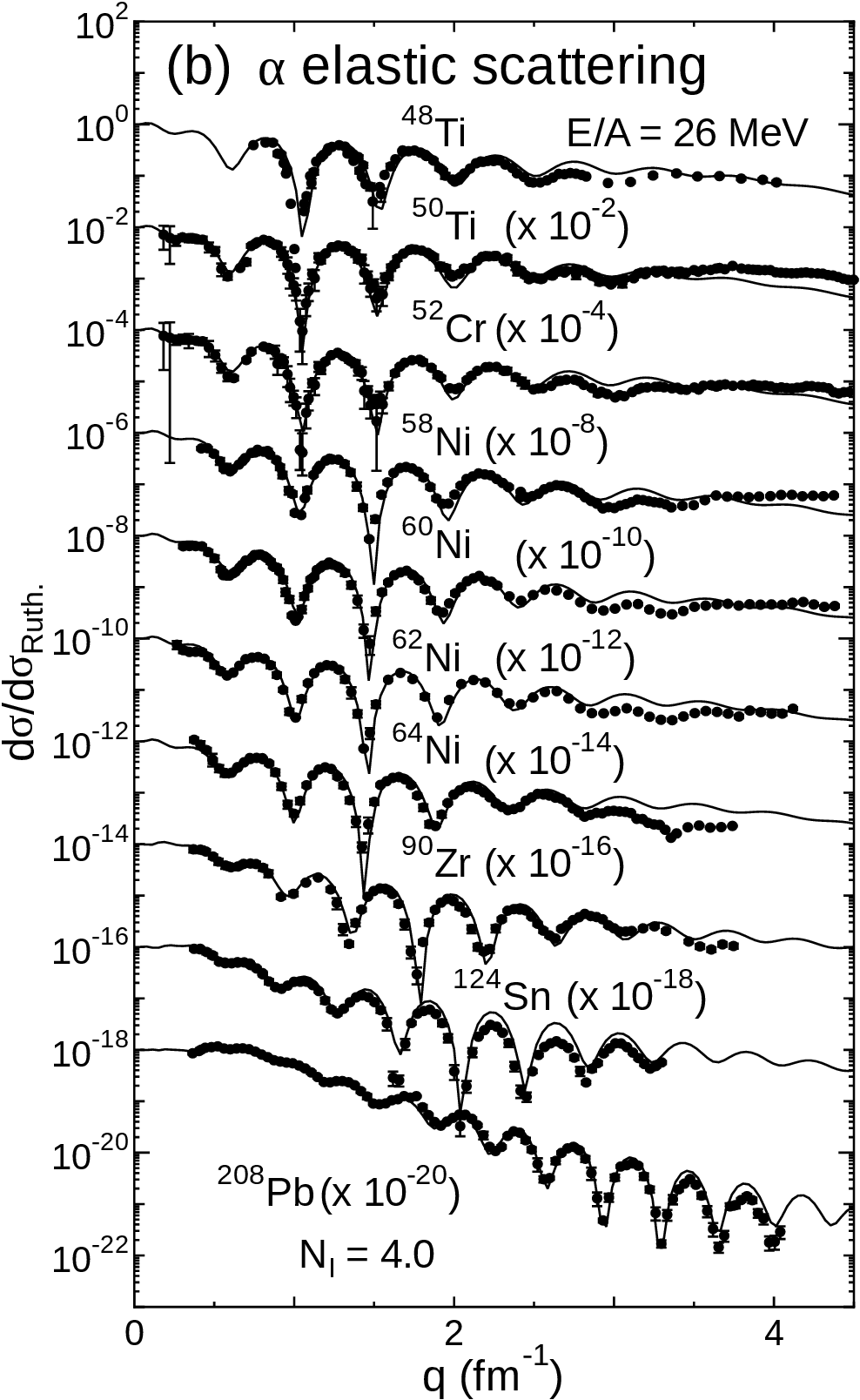}
\caption{Same as Fig.~\ref{fig:xs-low} but at $E/A =$ 26 MeV.
The experimental data are taken from Refs.~\cite{EXFOR,HAU69,RBE72,GIL80,REB74,PES83,REB72}.}
\label{fig:xs-mid}
\end{figure}
We also calculate the $\alpha$-nucleus elastic scattering cross sections at $E/A$ = 26 MeV.
Figure~\ref{fig:xs-mid} shows the cross sections for $A$ = 12--208 nuclei.
The experimental data are globally reproduced by the present DD-$\alpha N$ interaction.
Here, the $N_I$ values are fixed to be 4.0 which is roughly obtained from Fig.~\ref{fig:NI}.
For $^{12}$C and $^{16}$O targets, we adopt $N_I$ = 3.0 to reproduce the $\alpha$ elastic scattering.
The results indicate that the present DD-$\alpha N$ interaction works well to obtain a reasonable $\alpha$-nucleus potential.

\begin{figure}[ht]
\centering
\includegraphics[width=5cm]{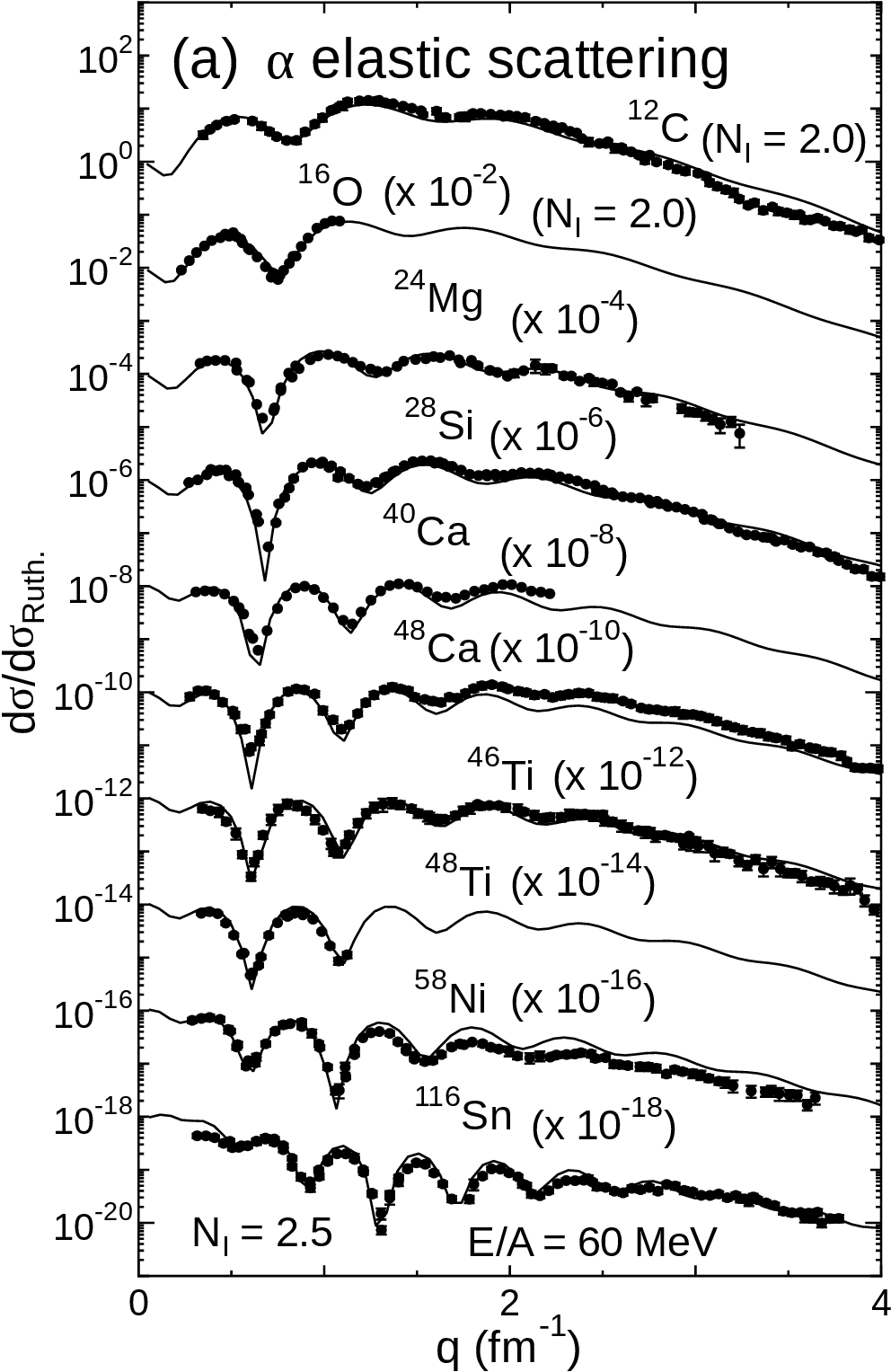}
\includegraphics[width=5cm]{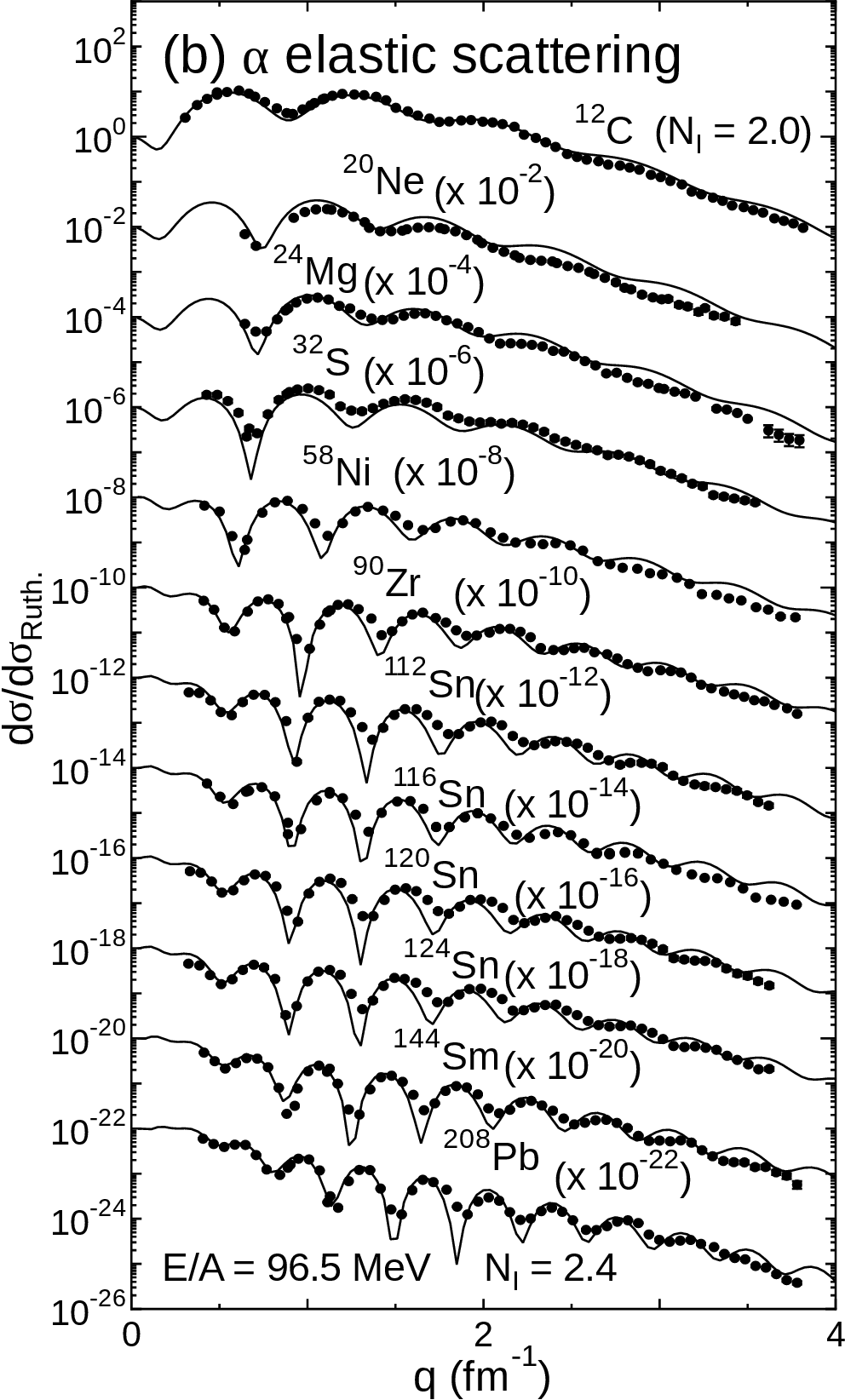}
\includegraphics[width=5cm]{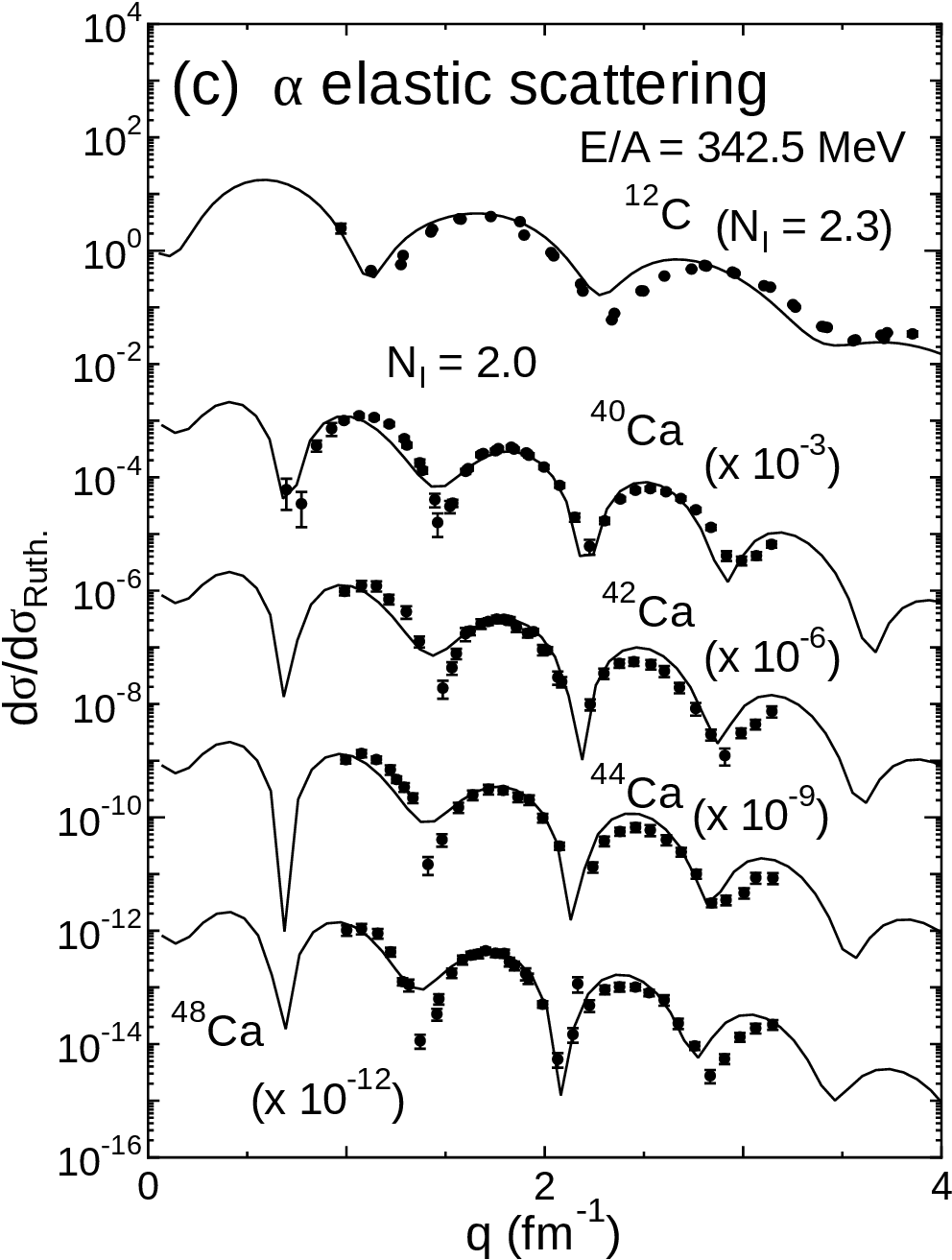}
\caption{Same as Fig.~\ref{fig:xs-low} but at $E/A =$ (a) 60, (b) 96.5, and (c) 342.5 MeV.
The experimental data are taken from Refs.~\cite{EXFOR,YOU98,LUI01,YOU99,YOU98-2,YOU97,LUI11,TOK06,LUI06,CLA98,ADA18,GUP15,ITO13,UCH04,LI10,ITO03,CHA76-2,ALK77}.}
\label{fig:xs-high}
\end{figure}

We also calculate the $\alpha$-nucleus elastic scattering cross sections at $E/A$ = 60, 96.5, and 342.5 MeV.
Figure~\ref{fig:xs-high} shows the results of $A$ = 12--208 nuclei.
Again, the experimental data are well reproduced by the present DD-$\alpha N$ interaction.
The $N_I$ values are drawn in Fig.~~\ref{fig:xs-high}, which are also roughly estimated from Fig.~\ref{fig:NI}.
Though we also need to readjust the strength of the $N_I$ value for the light target ($A \leq$ 16), we can obtain a reliable $\alpha$-nucleus potential with the present DD-$\alpha N$ interaction at $E/A \ge$ 60 MeV.

\section{Summary}
\label{sec:summary}

We have developed a global density-dependent (DD)-$\alpha N$ interaction to describe the $\alpha$-nucleus elastic scattering.
The present DD-$\alpha N$ interaction is constructed based on the phenomenological $p$ + $^4$He OMP, which describes the $p$ + $^{4}$He elastic scattering in a wide range of incident energies from 12.04--500 MeV per nucleon.
The real part of the $p$ + $^4$He potential is taken as a double Woods-Saxon form, which can accommodate drastic changes in the potential shape with increasing the incident energies.
The density-dependent part of the DD-$\alpha N$ interaction is phenomenologically fixed to reproduce the $\alpha$-nucleus elastic scattering cross-section data for selected target nuclei, $^{16}$O, $^{40}$Ca, $^{58}$Ni, $^{90}$Zr, and $^{208}$Pb.
The $\alpha$-nucleus elastic scattering cross-section data are well reproduced when the renormalization factor for the imaginary part of the $\alpha$-nucleus potential $N_I$ is properly chosen.
We show that the total reaction cross section is useful to fix the $N_I$ value.

To confirm the validity of the present DD-$\alpha N$ interaction, we demonstrate the $\alpha$-nucleus elastic scattering in the wide range of the target mass and the incident energy.
The $N_I$ parameter is roughly fixed with the reference to this paper.
The $\alpha$-nucleus elastic scattering is globally reproduced by the present DD-$\alpha N$ interaction.
When the present DD-$\alpha N$ interaction is applied to construct the $\alpha$-nucleus potential, we need a care as follows:
One parameter ($N_I$) is needed, but the value can be roughly read from Fig~\ref{fig:NI}.
At $E/A \sim$ 13 MeV and for $A \le$ 40, the backward angles may not be reproduced by the present DD-$\alpha N$ interaction.
For $A \le$ 16 targets, a reasonable $\alpha$-nucleus potential can be obtained with the readjustment of the $N_I$ value.
We have shown the applicability of the present DD-$\alpha N$ interaction to the $\alpha$-nucleus elastic scattering cross sections.
It is interesting to apply the present approach to the $\alpha$-nucleus inelastic scattering.
This application within the framework of the microscopic coupled-channel calculation is now in progress and will be reported in the forthcoming paper.

\section*{Acknowledgment}
The authors would like to acknowledge Professor Y. Iseri for providing the ALPS computer code to search the optical model potential.
This work was supported by the Japan Society for the Promotion of Science (JSPS) KAKENHI Grant Numbers JP18K03635, JP20K03943, JP20K03944, and JP22H01214.


\bibliographystyle{ptephy}

%
%
%
%
%

\end{document}